%% file: 0_main.tex
\documentclass[acmsmall, nonacm]{acmart}

\AtBeginDocument{%
  \providecommand\BibTeX{{%
    \normalfont B\kern-0.5em{\scshape i\kern-0.25em b}\kern-0.8em\TeX}}}

\acmJournal{TSAS}
\acmVolume{37}
\acmNumber{4}
\acmMonth{5}

\usepackage{subcaption}
\usepackage{booktabs}
\usepackage{multirow}
\usepackage{color, colortbl}
\usepackage{mathtools}
\usepackage{booktabs,caption}
\usepackage[flushleft]{threeparttable}
\begin{document}

\title{Responses to COVID-19 with Probabilistic Programming}

\author{Assem Zhunis}
\authornote{Both authors contributed equally to this research.}
\email{zhunis.assem@kaist.ac.kr}
\orcid{1234-5678-9012}
\author{Tung-Duong Mai}
\authornotemark[1]
\email{john_mai_2605@kaist.ac.kr}
\affiliation{%
  \institution{KAIST}
  \streetaddress{291 Daehak-ro}
  \city{Daejeon}
  \country{Republic of Korea}
  \postcode{34141}
}

\author{Sundong Kim}
\authornote{Supported by the
Institute for Basic Science (IBS-R029-C2).}
\affiliation{%
  \institution{Institute for Basic Science}
  \streetaddress{55 Expo-ro}
  \city{Daejeon}
  \country{Republic of Korea}
  \postcode{34126}}
\email{sundong@ibs.re.kr}


\renewcommand{\shortauthors}{Zhunis and Mai, et al.}

\begin{abstract}
The COVID-19 pandemic left its unique mark on the 21st century as one of the most significant disasters in history, triggering governments all over the world to respond with a wide range of interventions. However, these restrictions come with a substantial price tag. It is crucial for governments to form anti-virus strategies that balance the trade-off between protecting public health and minimizing the economic cost. This work proposes a probabilistic programming method to quantify the efficiency of major non-pharmaceutical interventions. We present a generative simulation model that accounts for the economic and human capital cost of adopting such strategies, and provide an end-to-end pipeline to simulate the virus spread and the incurred loss of various policy combinations. By investigating the national response in 10 countries covering four continents, we found that social distancing coupled with contact tracing is the most successful policy, reducing the virus transmission rate by 96\% along with a 98\% reduction in economic and human capital loss. Together with experimental results, we open-sourced a framework to test the efficacy of each policy combination.
\end{abstract}

\ccsdesc[500]{Theory of computation~Probabilistic computation}
\ccsdesc[500]{Social and professional topics~Governmental regulations}
\ccsdesc[500]{Applied computing~Health informatics}

\keywords{COVID-19, probabilistic programming, SEIRD model, simulation, economic impact, policy making}

\maketitle

\input{1_introduction}
\input{2_related_work}
\input{3_dataset}
\input{4_compartmental_model}
\input{5_policy_strength}
\input{6_simulation}
\input{7_conclusion}
\bibliographystyle{ACM-Reference-Format}
\bibliography{99_references}


\end{document}

%% file: 1_introduction.tex
\section{Introduction}

\label{sec:introduction}

The ongoing COVID-19 pandemic is one of the most challenging pandemics in human history, infecting more than 170 million people worldwide with more than 3.5 million fatalities as of May 30, 2021 \cite{dong2020interactive}. Rapid and easy transmission of COVID-19 leads to a high and fast-growing caseload,  overwhelmingly straining the healthcare systems of many countries. Governments are pushed to apply prompt and effective interventions to protect public health. Such policies include lockdown, social distancing, contact tracing, hygiene and mask mandates. However, countries differ on these measures and their stringency due to differences in public acceptance, the political climate, or government priority. Thus, many interventions were applied considering the individual socioeconomic status of countries. Furthermore, most countries lacked experience in handling the pandemic, only a handful have successfully brought the pandemic under control. The world has witnessed how the initial response to the virus dictated the trajectory of the virus spread. 

\begin{figure}[t]
\begin{center}
    \includegraphics[width=0.55\textwidth]{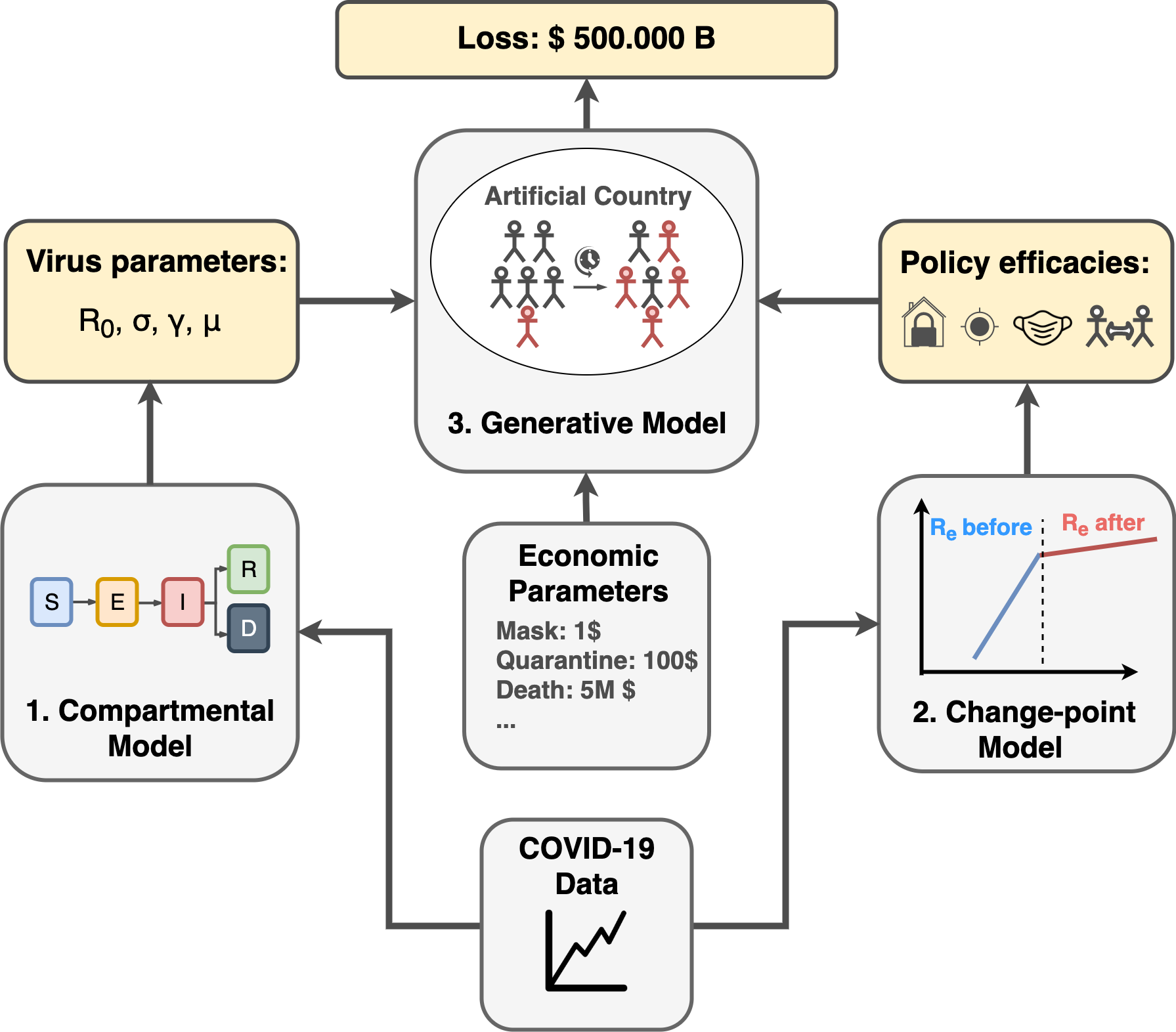}
   \caption{Project pipeline. First, we infer COVID-19 related parameters such as basic reproduction number $R_0$, incubation rate $\sigma$, recovery rate $\gamma$, and mortality rate $\mu$ using the compartmental model. Second, we apply the change-point model to infer policy efficiencies from different countries. Finally, using inferred parameters from previous steps and economic parameters from real-world data, we run the generative model in artificial country simulation to estimate the economic cost for different policy combinations.
   }

    \label{pipeline}

\end{center}

\end{figure}

Apart from its health impact, coronavirus has affected the economic state of the world with various restrictions imposed by governments to mitigate the virus spread. The pandemic already caused a bigger recession than the Great Depression~\cite{wheelock2020comparing}. On April 18, 2020, in 158 out of 181 countries, workplaces were temporarily closed, which hurts companies in the absence of essential workers due to multiple waves of the pandemic~\cite{hale2020variation}. As reported by Mandel et al., lockdown generates more than 33\% drop in global output at its peak and more than 9\% drop in annual GDP \cite{mandel2020economic}. Furthermore, adverse economic effects of lockdown could even diffuse to the neighboring countries by supply chains~\cite{inoue2020propagation}. Thus, governments should carefully take economic context into account when making policy decisions.

This paper proposes a probabilistic programming method to evaluate the strategies imposed by different countries and point out which policies are the most successful in the initial response to the crisis. To provide insights on the effectiveness of initial responses to the pandemic, we analyzed data from 10 countries covering four continents. Moreover, we present a method to balance economic trade-offs of adopting specific policies by providing a generative model that considers the economic context of a given country. Given the recent focus on vaccination efforts, we also examine the effect of vaccination in the containment of the coronavirus in Israel and the United States.

To quantitatively express and analyze the success and failure of different countries, we utilize a probabilistic approach to tackle the COVID-19 transmission dynamics. 
As illustrated in Figure~\ref{pipeline}, our approach has three major components: 

 \begin{enumerate}
    \item Infer COVID-19 statistics by the compartmental model (\S~\ref{sec:compartmental}).
    \item Estimate policy strengths by the change-point model (\S~\ref{sec:policy}).
    \item Simulate virus in the context of policy combinations considering the economic loss by the generative model. (\S~\ref{sec:simulation}).
\end{enumerate}

The compartmental model is to understand the representative statistics of the virus transmission dynamics, including recovery time, incubation time, reproduction number ($R_0$), and mortality rate. We infer the baseline statistics by fitting the SEIRD compartmental model on the Swedish data before the Swedish government imposed any policies. We assume that these statistics represent the original virus features unaffected by any human interventions. The obtained statistics serve as the default parameters for later modeling parts.


With the change-point model, we estimate the strength of the policies applied to curb the virus spread. Countries around the world impose various interventions with different degrees. 
Furthermore, the populations worldwide are largely not homogeneous; therefore, the same policy could have different outcomes in different populations. Instead of capturing all these complicated factors, we choose cases where a particular policy can viably represent the upper bound of these policy's efficiency, i.e., the maximum reduction in infection rate. This provides a good idea of how effective each policy would be if applied in full force. To find these upper bounds, we investigate the countries with a successful initial response to the pandemic that stringently applied a given measure, such as China for lockdown or Singapore for social distancing. As these countries curbed the first wave of the spread by firmly applying a particular measure, we can consider the effect in these countries as the maximum effect the measure could perform. The measure will introduce an abrupt change in caseload growth with a significant drop in growth rate, starting from a \emph{change point} in the timeline. We utilize the growth rate before and after this \textit{change-point} to detect the effect of each measure. We run several experiments on countries with different initial responses to the pandemic. The efficiency of the initial policies is represented in terms of the transmission rate change after the policy establishment. Inference result suggests that all major interventions are effective in reducing the virus spread. For example, contact tracing coupled with social distancing yields 96\% reduction in the virus transmission rate, achieving the same effect as lockdown and outperforming all other policies.

The last part of our study includes simulating the virus in an imaginary country that follows all virus and policy statistics inferred from the previous parts. Our generative model is to support decision-makers to solve optimization problems having opposing objectives: public health and the economy. Stringent measures indeed incur economic collapse, but loosening the measures could lead to a devastating crisis. Therefore, the trade-off should be considered carefully. Our model predicts the trajectory of the pandemic, including cases, deaths, and recoveries. Moreover, we incorporate the economic cost into the simulation to address the economic trade-off of policy establishments. By controlling parameters, we estimate how the pandemic plays out in different scenarios and conclude which policy combination can effectively mitigate the virus in public health and economic dimensions. Simulation results suggest contact tracing coupled with social distancing incurs the lowest economic and human capital loss.



With all these analyses, we provide a simple but insightful model to analyze several features of a pandemic: severity of the disease, policy efficiency, and economic impact. This will help to understand the success and failure of each country in its response to the pandemic. It could be used as a playbook to to better prepare for a possible pandemic in the future. For reproducibility, the code and datasets used in the paper are available at \url{https://git.io/JGcPW}.

The rest of this paper is organized as follows. In Section~\ref{sec:related}, we review related work. In Section~\ref{sec:dataset}, we introduce the dataset. In Sections~\ref{sec:compartmental} and \ref{sec:policy}, we propose our compartmental model and estimate policy strength by change-point model, respectively. With this model and economic viewpoint, we simulate an artificial country by changing policies in Section~\ref{sec:simulation}. We conclude the paper in Section~\ref{sec:conclusion}.

%% file: 2_related_work.tex
\section{Related Work}
\label{sec:related}

\subsection{Compartmental models}
\label{sec:related:compartmental-models}

Most of the epidemic models divide the target population into a certain number of compartments, consisting of individuals with identical statuses concerning a given disease. The foundations of the entire approach to epidemiology based on compartmental models were laid by public health physicians in the early 1900s.  One of the first applications of the compartmental model was made by R. Ross, who demonstrated the dynamics of the transmission of malaria between mosquitoes and humans and consequently was awarded the Nobel Prize in Medicine in 1902~\cite{rajakumar1999centennial, ross1911some}. Since then, compartmental models are still widely used to simulate the spread of a variety of infections~\cite{Brauer2008}. 

One of the most popular extensions of the SIR model is the SEIR model~\cite{li1995global}, a traditional method used to simulate infectious disease that incubates inside the hosts for a while before the hosts become infectious. The SEIR model considers the incubation period by introducing a new compartment $E$ (Exposed) to the compartmental system. This model and its modifications were already adapted to simulate the COVID-19 virus in many countries~\cite{ANNAS2020110072, pandey2020seir, yang2020modified}. In this work, we use a modified version of the model---SEIRD~\cite{piovella_2020} with the death compartment $D$. 

\subsection{Probabilistic algorithms} 
\label{sec:related:probabilistic-algorithms}


The Markov chain Monte Carlo~(MCMC) is a large class of sampling algorithms widely used for probabilistic problems. MCMC was first introduced in 1953 as a new method to simulate the distribution of states for the system of idealized molecules~\cite{metropolis1953equation}. However, the application of the algorithm did not limit itself to the physics field. It was later adapted and generalized by Hastings~\cite{hastings1970monte} to focus on statistical problems, opening its application to a wide range of applications. Due to its ability to handle complex types of analyses, the MCMC approach was widely used in finance~\cite{eraker2001mcmc,johannes2010mcmc}, communication~\cite{farhang2006markov, al1995stochastic}, computational biology~\cite{gupta2014comparison}, linguistics~\cite{gill2008partial, wells2004generalized} and other fields with probabilistic settings. By no surprise, these methods are widely popular for estimating effects in complex epidemiological analyses as well~\cite{hamra2013markov,lawson1995mcmc}. For example, Cauchemez et al. has shown how to model influenza transmission using the Bayesian MCMC approach~\cite{cauchemez2004bayesian}, and lots of variations of MCMC methods were used to infer features of an Ebola virus and analyze its transmission mechanism~\cite{lekone2006statistical, ndanguza2013statistical}. Recent reports are also benefited from the Bayesian MCMC methods to infer COVID-19 virus transmission dynamics. Zhou et al. implemented such inference based on a probabilistic compartmental model using daily confirmed COVID-19 cases and applied it to six states of the United States \cite{zhou2020semiparametric}.

MCMC algorithms are also successfully applied to change-point models. The objective is to detect the abrupt property changes lying behind the time-series data~\cite{brooks1998markov}. Recent work showed that MCMC algorithms with Bayesian parameter inference could be used to detect change-points in COVID-19 spread using SIR and SEIR epidemiological models of South Africa~\cite{mbuvha2020bayesian}. According to their results, South Africa experienced two change-points: the first at the time of the national lockdown and the second after the massive screening and testing program. Dehning adopted a similar approach, Jonas et al., to the case study of coronavirus spread in Germany by utilizing SIR models with MCMC sampling for detecting change-points in effective growth rate that correlates well with the times of publicly announced interventions~\cite{dehning2020inferring}.
Following these examples, we applied the change point model with an extended SEIRD epidemiology model to identify where policies affected the COVID-19 virus transmission rate in 10 countries with different interventions.

\subsection{Policy strength estimation}
\label{sec:related:policy-strength-estimation}


A wide range of work was done to estimate the efficiencies of the policies imposed by different countries to prevent COVID-19. Many of them were focused on the individual country cases considering their unique demographic features~\cite{kavaliunas2020swedish,debnath2020india,naumann2020covid}, while other reports compared many countries by the independent effects of a single category of policy~\cite{chinazzi2020effect,arenas2020derivation,teslya2020impact}. For instance, Iwata et al. used Bayesian method analysis~\cite{iwata2020school}. They did not reveal the effectiveness of school closures that occurred in Japan in mitigating the risk of coronavirus infection in the nation. Another recent work by Sharov et al. used a modified SIR model to compare the effectiveness of lockdown measures introduced during the coronavirus pandemic in 13 European countries, comparing them to two baseline countries (Sweden and Iceland) that did not implement the lockdown policies. For evaluation, this work used the herd immunity level and time of formation to indicate the effectiveness of lockdown measures~\cite{sharov2020creating}. According to Sharov's results, lockdown and no-lockdown modes of containment led to roughly similar results. 

There are also reports considering multiple policies across the globe~\cite{hale2020variation,jinjarak2020accounting}. One example is work by Flaxman, S., Mishra, S., Gandy, A. et al. ~\cite{flaxman2020estimating}, which investigates effects of applied non-pharmaceutical interventions (NPIs) across 11 European countries for the period from the start of the COVID-19 epidemics. According to their results, major non-pharmaceutical interventions, like lockdowns, have had a significant effect on reducing transmission of the virus.  However, a subsequent study by Haug et al.~\cite{haug2020ranking}, which assessed the efficacy of 6,068 NPIs across 226 countries and gave a detailed analysis of the country-specific 'what-if' scenarios, showed different results. They analyzed the impact of government interventions on the effective reproduction number $R_t$ by combining several analytical approaches. By utilizing statistical, inference, and artificial intelligence tools, they concluded that combinations of some less disruptive and less costly NPIs could be as effective as more expensive and harsh ones like national lockdowns.  Brauner et al. came to the same conclusion by analyzing 41 countries during the first wave of the pandemic. According to their study, less harsh NPIs can be more effective in mitigating COVID-19 transmission than more strict stay-at-home orders\cite{brauner2021inferring}.

However, one of the significant limitations of recent studies is that none of them perform a comprehensive analysis considering the economic factors that affect the efficiencies of the policies. There were some reports regarding the economic cost of the pandemic situation across the globe~\cite{maital2020global}. For example, McKibbin and Fernando et al. simulated a global economic model to explore seven scenarios, which differ in the proportion of the population who become infected or dead~\cite{mckibbin2020economic}. According to their estimations, in a scenario where COVID-19 develops into a global pandemic, the cost of lost economic output begins to escalate into trillions of dollars~\cite{mckibbin2020global}. However, they do not include the effect of policy interventions in their simulations. To address this limitation, we propose another method of cost estimations by involving policy effects in \S~\ref{sec:simulation}.

%% file: 3_dataset.tex
\begin{figure}[htbp]
    \centering
    \begin{subfigure}[b]{0.32\linewidth}
    \includegraphics[width=\linewidth]{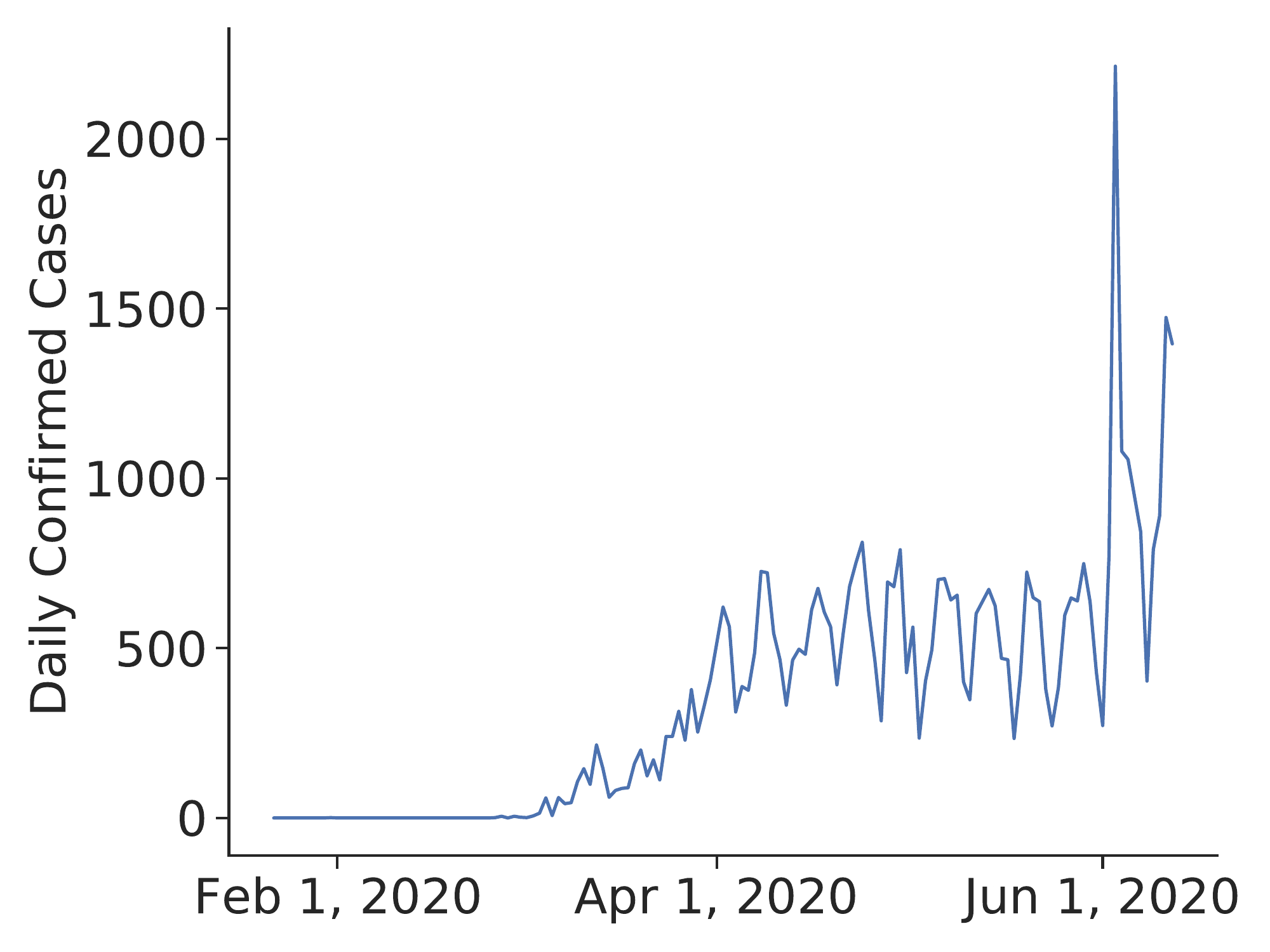}
    \caption{Sweden}
    \end{subfigure}
    \begin{subfigure}[b]{0.32\columnwidth}
    \includegraphics[width=\linewidth]{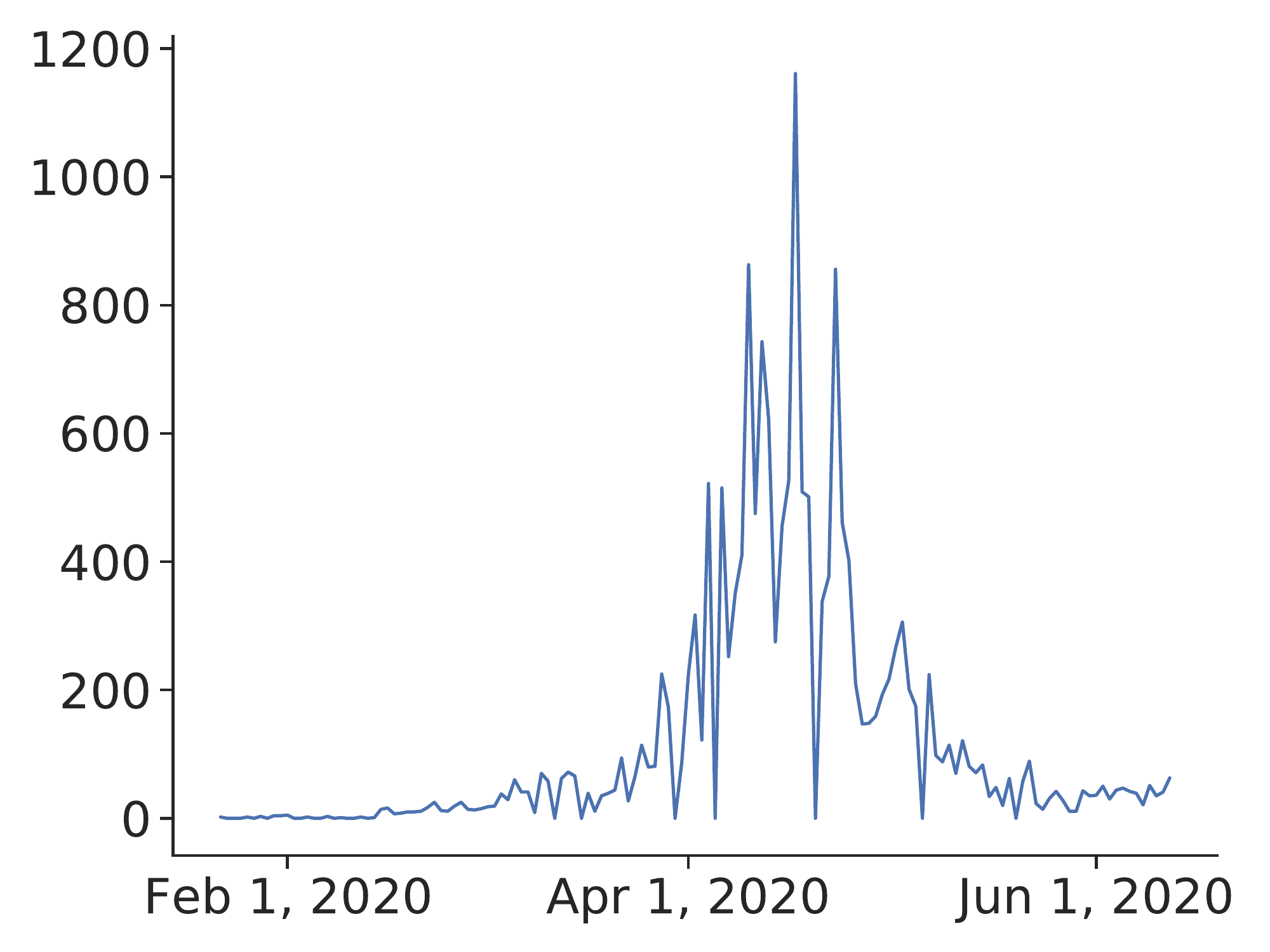}
    \caption{Japan}
    \end{subfigure}
    \begin{subfigure}[b]{0.32\linewidth}
    \includegraphics[width=\linewidth]{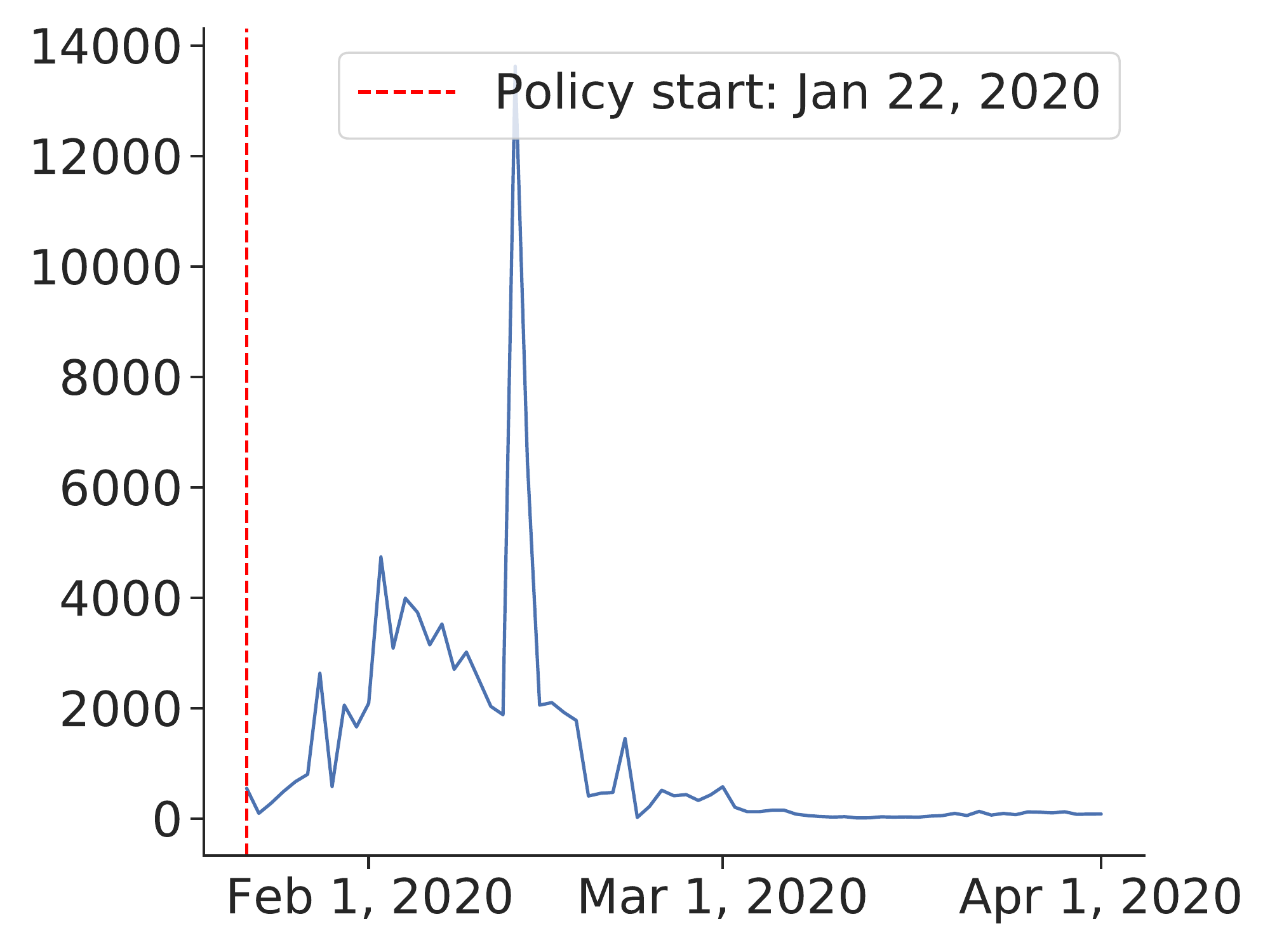}
    \caption{China}
    \end{subfigure}
    \begin{subfigure}[b]{0.32\linewidth}
    \includegraphics[width=\linewidth]{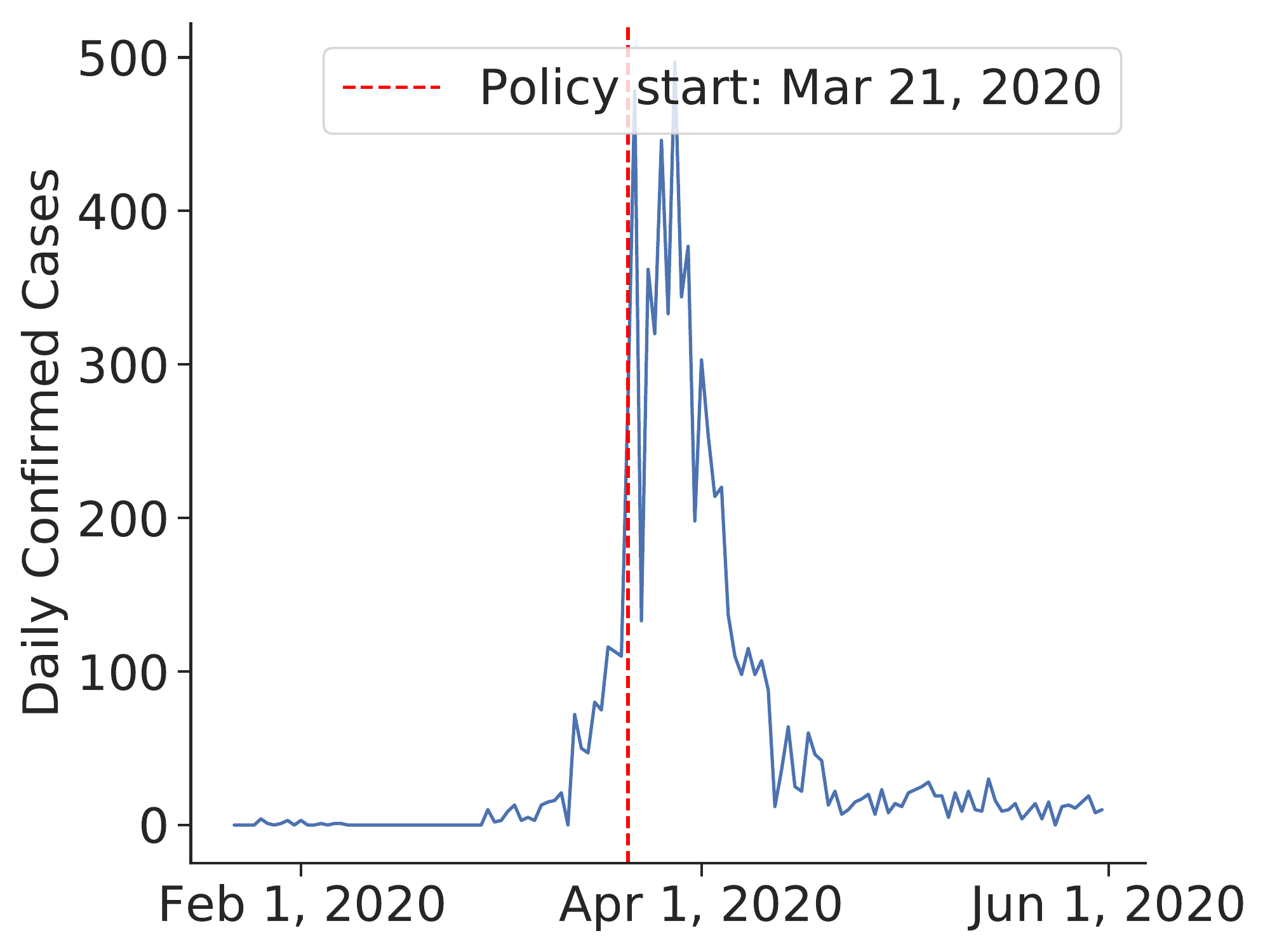}
    \caption{Australia}
    \end{subfigure}
    \begin{subfigure}[b]{0.32\linewidth}
    \includegraphics[width=\linewidth]{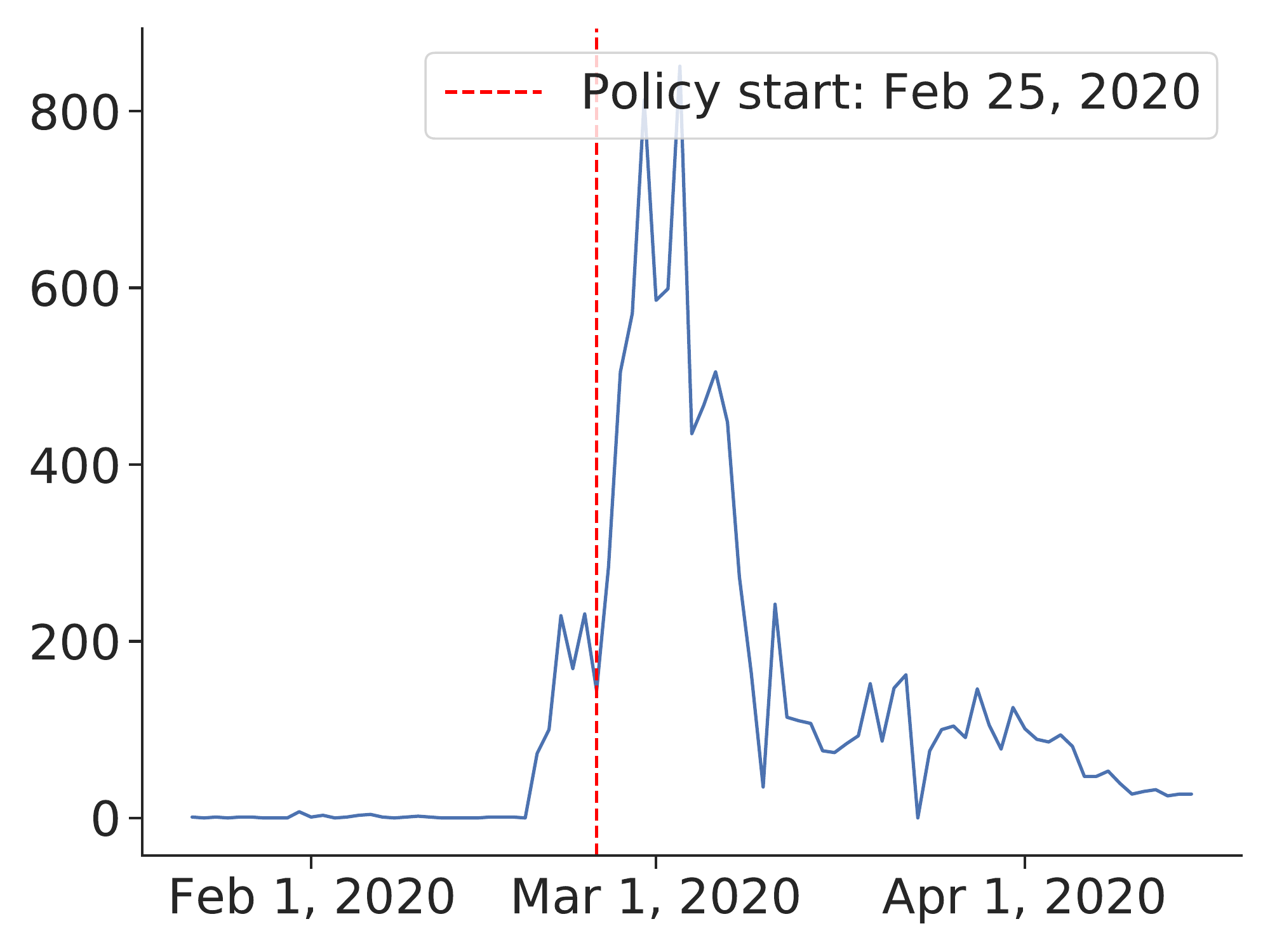}
    \caption{Korea}
    \end{subfigure}
    \begin{subfigure}[b]{0.32\linewidth}
    \includegraphics[width=\linewidth]{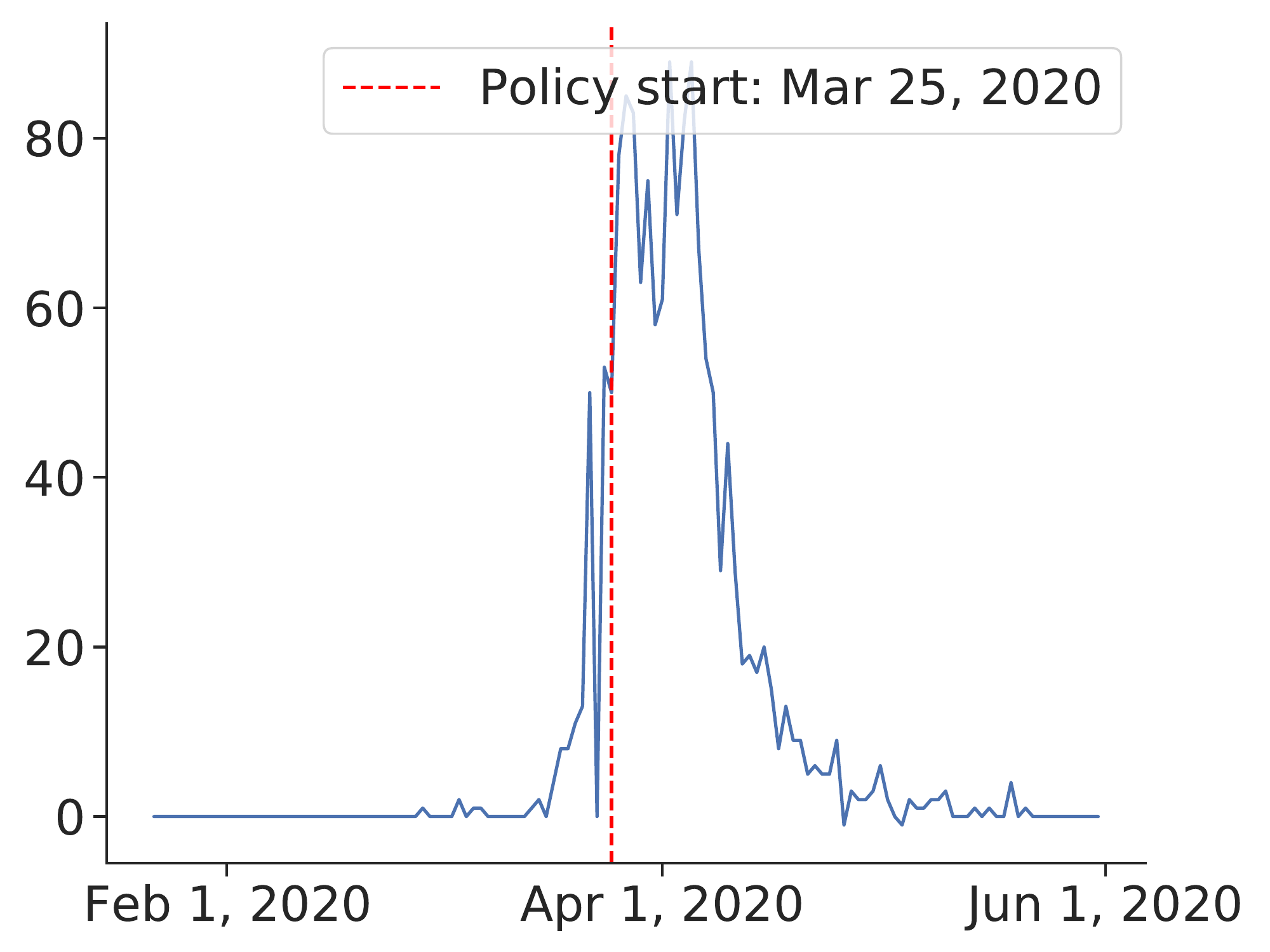}
    \caption{New Zealand}
    \end{subfigure}
    \begin{subfigure}[b]{0.32\linewidth}
    \includegraphics[width=\linewidth]{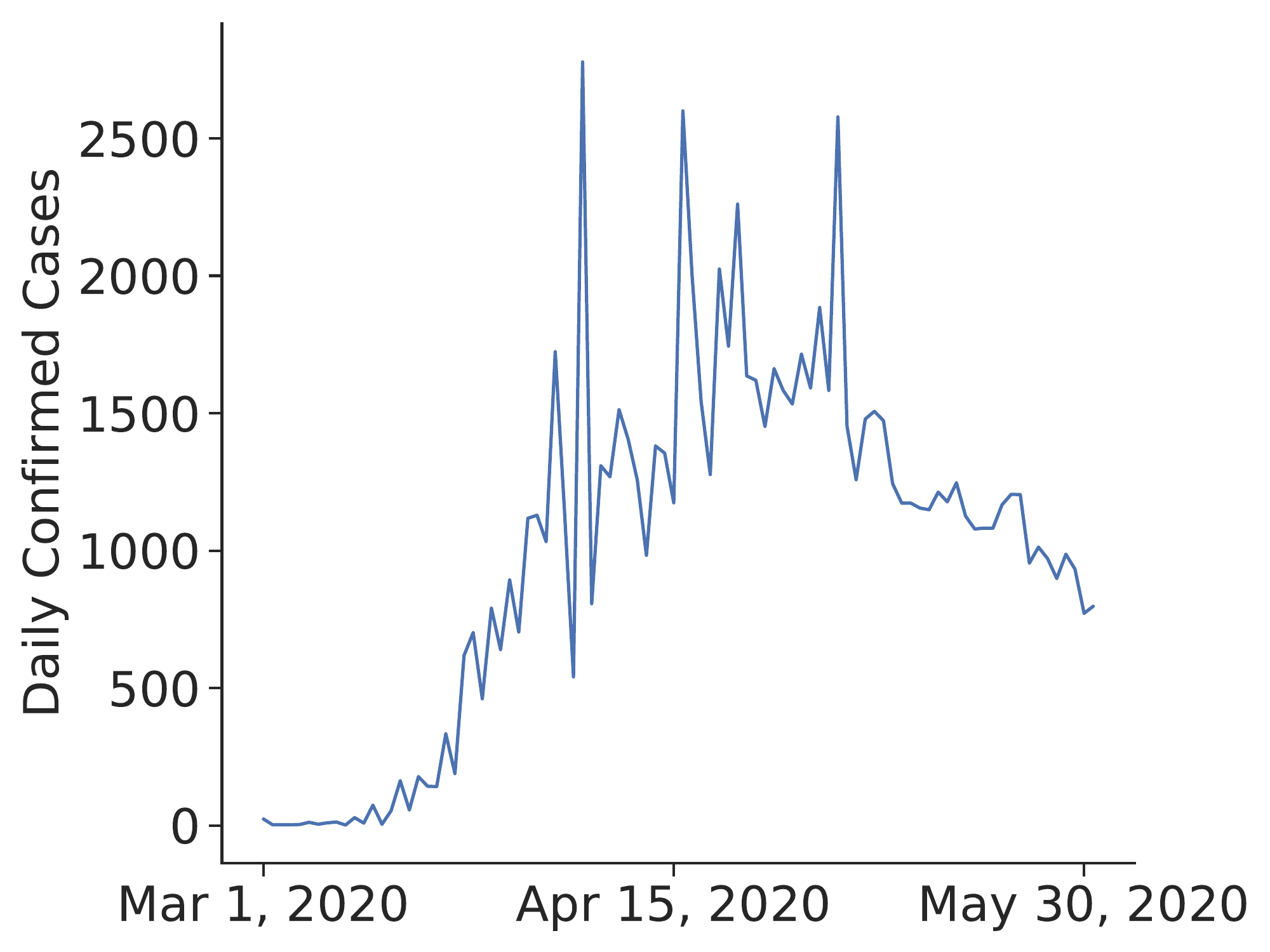}
    \caption{Canada}
    \end{subfigure}
    \begin{subfigure}[b]{0.32\linewidth}
    \includegraphics[width=\linewidth]{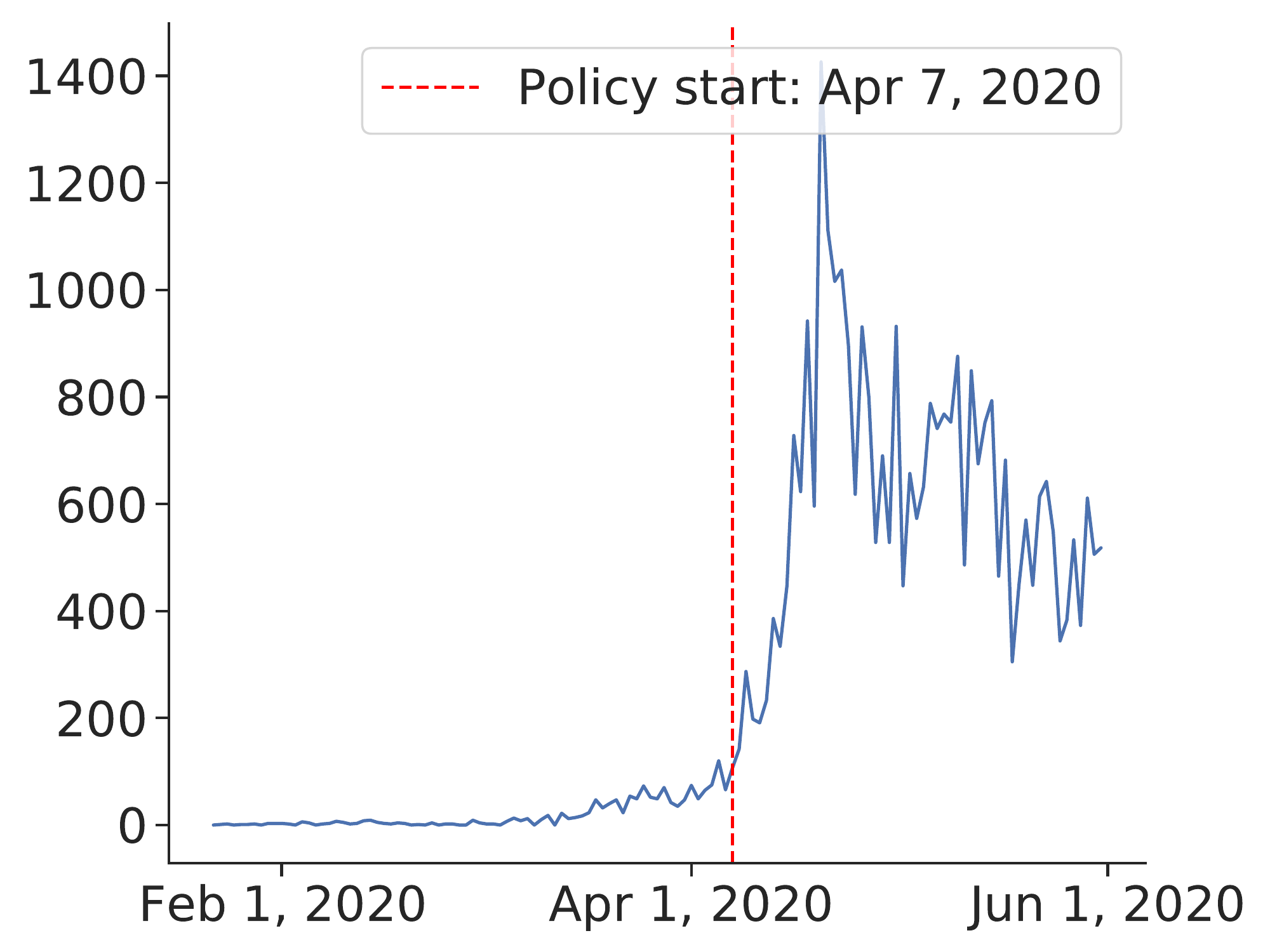}
    \caption{Singapore}
    \end{subfigure}
    \hfill
    \caption{Daily number of confirmed cases of 8 countries. Vertical red lines indicate the date when a certain policy was established. In the case of Sweden and Japan, we consider the time period before the policy establishment. For Canada, the policy was established gradually. Thus, the exact dates of policy start for this country are not determined.
    }
    \label{daily-confirmed-cases}
\end{figure}

\begin{figure}[htbp]
    \centering
    \begin{subfigure}[b]{0.45\linewidth}
    \includegraphics[width=\linewidth]{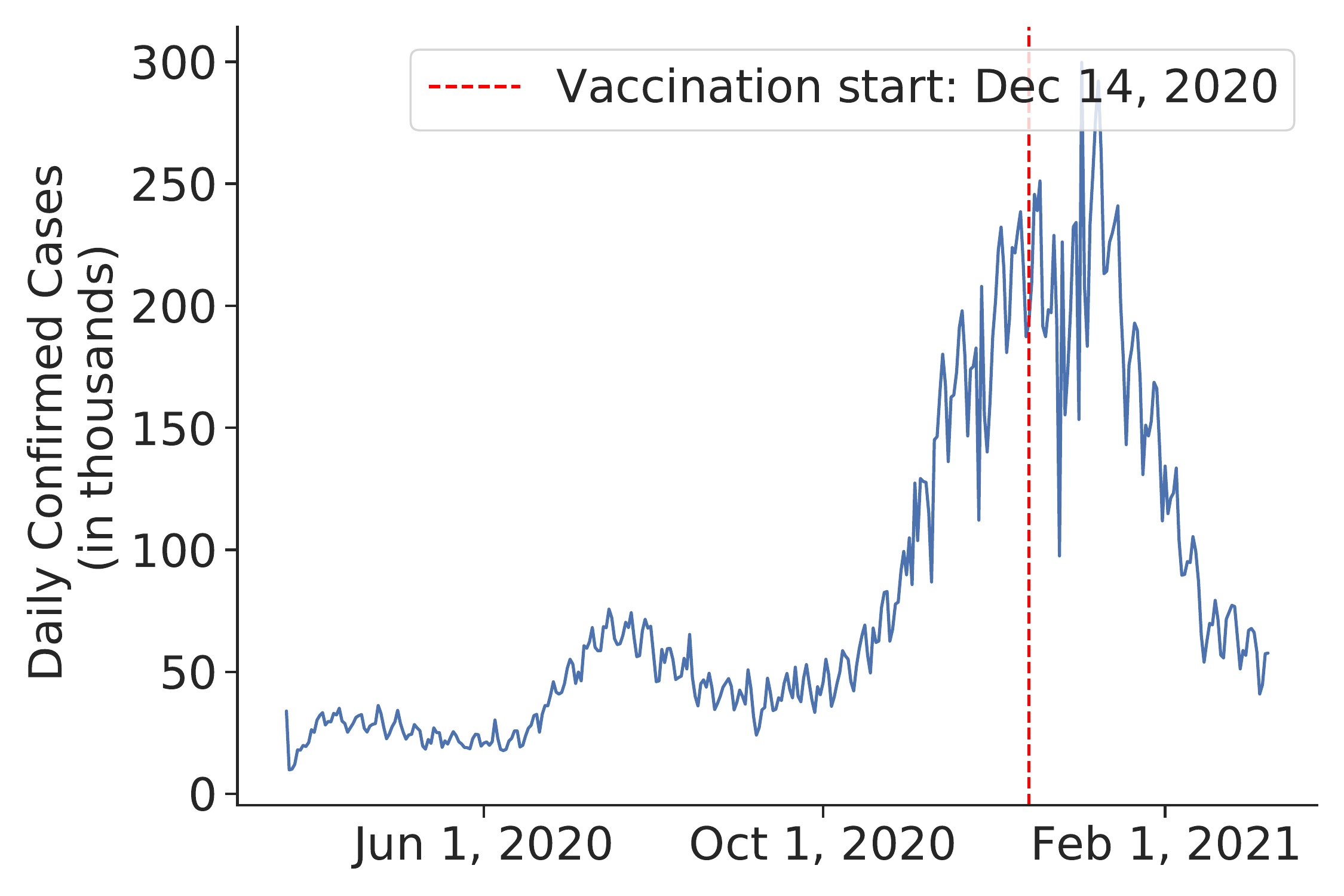}
    \caption{United states}
    \end{subfigure}
    \begin{subfigure}[b]{0.45\linewidth}
    \includegraphics[width=\linewidth]{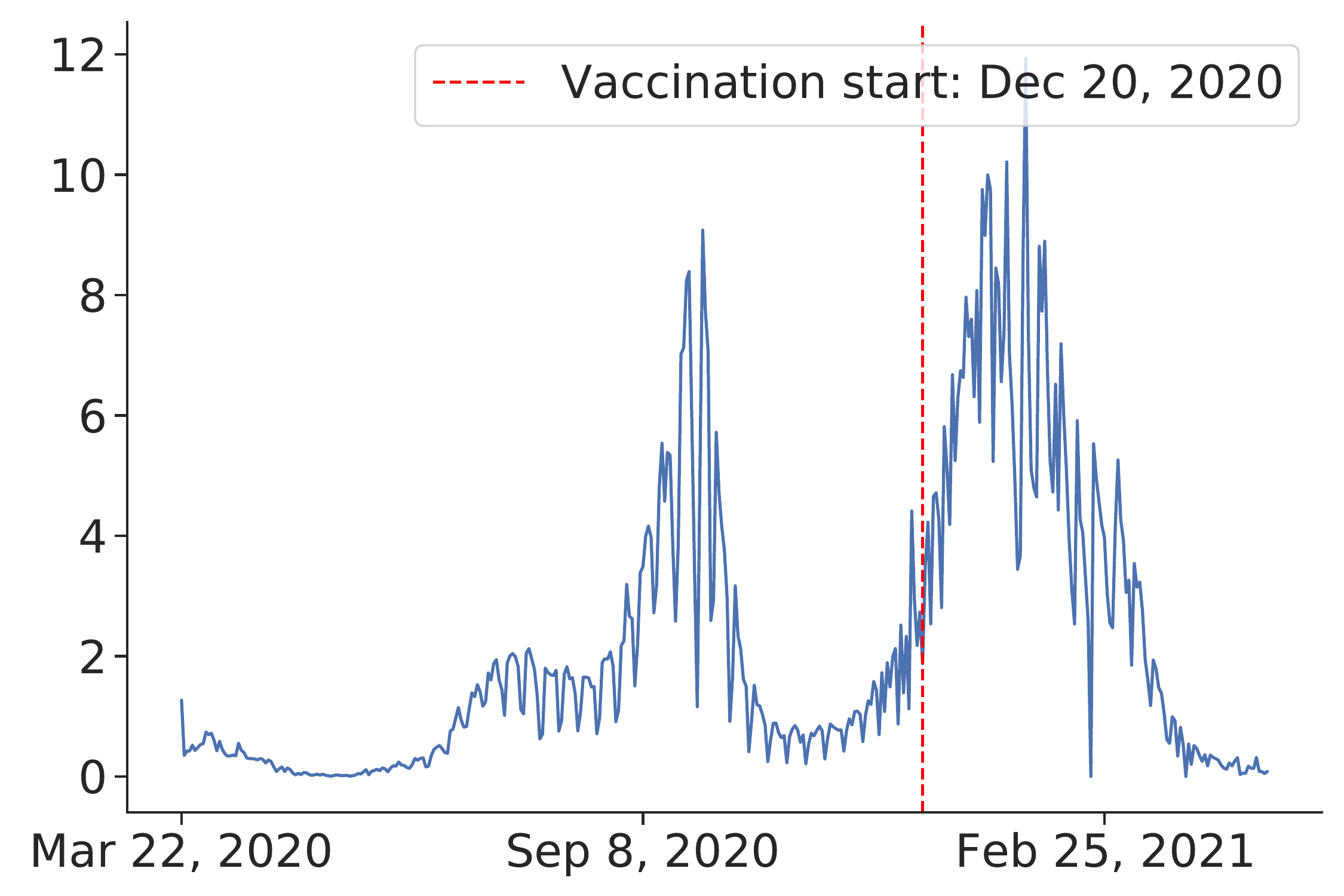}\caption{Israel}
    \end{subfigure}
    \hfill
    \caption{Daily number of confirmed cases of the countries in a timeline in which we investigate the effect of vaccines.}
    
    \label{data}
\end{figure}

\section{Dataset}
\label{sec:dataset}
For our analysis, we used virus data from two sources that are available online. The first dataset is taken from Kaggle. 
This COVID-19-related dataset\footnote{\url{https://www.kaggle.com/imdevskp/corona-virus-report?select=full\_grouped.csv}} was collected from the John Hopkins  University dashboard\footnote{\url{https://coronavirus.jhu.edu/map.html}} and Worldometers website\footnote{\url{https://www.worldometers.info/coronavirus/}}. The data for Israel and the US was taken from the COVID-19 Data Repository by the Center for Systems Science and Engineering (CSSE) at Johns Hopkins University\footnote{\url{https://git.io/Jvoxz}}. Both datasets report the number of confirmed, death, and recovered cases for each day since the first confirmed case across the globe, divided by countries, regions, and provinces. The plots of daily confirmed cases for 10 countries used for analysis can be found in Figures ~\ref{daily-confirmed-cases} and ~\ref{data}.

%% file: 4_compartmental_model.tex
\section{Analyze COVID-19 Statistics by Compartmental Model}
\label{sec:compartmental}

In this section, we will introduce the proposed methodology to infer SARS-CoV-2 virus statistics. We will first present the SEIRD epidemiological compartmental model and the corresponding probabilistic programming model to infer several virus statistics. Then, we perform inference on the Sweden data to infer the important virus statistics.

\subsection{Compartmental SEIRD Model}
\label{sub:seird}
The most basic compartmental model, namely the SIR model, uses three compartments of Susceptible (S), Infectious (I), and Recovered (R). Each individual can move from a compartment to another compartment, resembling the progress of the disease. We could use S, I and R to denote the number of individuals in their respective compartments. For the COVID-19 case, there is an incubation period in which people are infected but not yet infectious. Hence, we adapted the epidemiological SEIRD model~\cite{piovella_2020} for our simulations, extending the SIR model with the E compartment of exposed individuals and the D compartment for deaths. 
\begin{itemize}
    \item Susceptible (S): Individuals in the compartment are neither infected nor immune to the diseases and, hence, could contract the disease. If the susceptible individuals contract the disease (via contact with an infectious individual), they progress to the Exposed compartment.
    \item Exposed (E): Individuals in the compartment are infected but unable to pass the disease to susceptible individuals. If the Exposed individuals finish their incubation period and can infect others, they progress to the Infectious compartment.
    \item Infected (I): Individuals in the compartment are infected and pass the disease to susceptible individuals. If the Infectious individuals recover from the disease and carry immunity or die from the disease, they progress to the Recovered (or Resistant) and Dead compartments, respectively.
    \item Recovered (R): Individuals in the compartment are immune to the disease. If the Recovered individuals lose their immunity, they progress to the Susceptible compartment.
    \item Dead (D):
    Individuals in the compartment cannot progress to any other compartment.
\end{itemize}

The below differential equations (Eq.~\ref{eq:SEIR_S}--\ref{eq:SEIR_D}) describe the transition between compartments~\cite{piovella_2020}.
\begin{align}
\frac{dS}{dt}&=-\frac{R_e \gamma S}{N} I + \alpha R, \label{eq:SEIR_S} \\
\frac{dE}{dt}&=\frac{R_e \gamma S}{N} I - \sigma E, \label{eq:SEIR_E} \\
\frac{dI}{dt}&=\sigma E- \gamma I, \label{eq:SEIR_I} \\
\frac{dR}{dt}&=\gamma (1- \mu)I- \alpha R, \label{eq:SEIR_R} \\
\frac{dD}{dt}&= \gamma \mu I,\label{eq:SEIR_D}\\
N &= S+E+I+R+D, \label{eq:SEIR_N}
\end{align}
where the effective reproduction number ($R_e$) is the expected number of people that each infected individual can transmit the virus to during the outbreak, the basic reproduction number ($R_0$) is the natural reproduction number when there is no intervention, the incubation time ($t_E = \frac{1}{\sigma}$) is the average time in which an individual is exposed but not yet infectious, the recovery time ($t_I = \frac{1}{\gamma}$) is the average time after which an the infected case become concluded (recovered/dead), the case fatality proportion $\mu$ is the proportion of fatal cases among all concluded cases, and the waning time ($t_R = \frac{1}{\alpha}$) is the time that recovered individuals retain immunity. In this work, we assume people carry lifelong immunity to the disease upon recovery ($\alpha = 0$), and the population stays constant over time and is equal to N.

\subsection{Scaling Up with Probabilistic Programming} 
\label{subsub:mcmc}
To implement the probabilistic models, we used the probabilistic programming language Pyro~\cite{bingham2018pyro}. For this particular inference task, we adopted Pyro’s Epidemiology framework~\cite{epidemiology_pyro_documentation} for scaling up our experiments with a restricted class of stochastic discrete-time discrete-count compartmental models. This framework uses the MCMC algorithm to fit the SEIRD model to infer COVID-19-related parameters: reproduction number $R_0$, recovery time, incubation time, transmission rate, and mortality rate. 



MCMC is a stochastic algorithm that repeatedly generates random samples describing the distribution of parameters of interest (in our case, COVID-19 related parameters), where a new sample is generated based on the previous sample, thereby creating a Markov chain. The Markov chain has a stationary probability $p_S(x)$ such that if the chain ever arrives at $p_S(x)$, it will keep sampling from $p_S(x)$ forever. Therefore, the goal of MCMC is to design a transition probability to make the stationary distribution equate the target probability (i.e. $p_S(x) = p(x)$). Starting from an initial random sample, the algorithm guides the Markov chain to the stationary distribution, which we force to be the same as our target distribution \cite{andrieu2003introduction}.


A popular instance of the MCMC method is the Metropolis-Hastings algorithm that uses sampled proposal probability distribution (also called the kernel), followed by an acceptance criterion that chooses to accept or discard the new sample by comparing how likely the proposal distribution is to differ from the true next-state probability distribution.
For optimizing the sampling process, we used an instance of the Metropolis-Hastings algorithm, namely the Hamiltonian Monte Carlo (HMC) algorithm with the No-U-Turn Sampler (NUTS). 
 The HMC algorithm avoids random walk behavior by taking steps informed by the first-order gradient information~\cite{nuts}. It utilizes an approximate Hamiltonian dynamics simulation, which is then corrected by a Metropolis acceptance step \cite{betancourt2013hamiltonian, neal2012mcmc}. HMC reduces the correlation between successive sampled states, allowing the algorithm to converge much faster with fewer Markov chain samples. However, since HMC is highly sensitive to two hyper-parameters: step size $\alpha$ and the number of steps $L$, the No-U-Turn Sampler (NUTS) is used to adaptively set these parameters~\cite{nuts}. Thus, we can perform HMC without any manual hyperparameter tuning.

\subsection{Fitting SEIRD Model to Sweden -- The Reference Country} 
We ran the model on the Swedish data before April 1st, 2020. We chose this early stage of the COVID-19 pandemic since Sweden did not impose any strict policies and aimed to achieve herd immunity \cite{sweden}. We assumed that the virus transmission rate was unaffected by any interventions, so we used Sweden as a baseline case and perform the experiments to infer unaffected COVID-19 parameters. To run our probabilistic model, we set the prior according to the estimations of the World Health Organization \cite{world_health_organization}. It was reported that mild cases typically recovered within two weeks, the incubation period was on average 5–6 days, and $R_0$ was typically around 2. Mortality and recovery rates differed depending on the region and stage of the virus spread, but the case-fatality rate was roughly 2.5\%. \looseness=-1

The obtained posterior values for Swedish data are shown in Table 1. For more accurate results, we ran the model 6 times and reported the averaged values. The results are reasonable enough to use in our further simulations.

\begin{table}[ht]
\begin{center}
\begin{tabular}{l c r} 
\toprule
\textbf{Parameter} & \textbf{Abbreviation}& \textbf{Value}\\
\midrule
Recovery time & $1/\gamma$ &  16.33 days \\
Incubation time & $1/\sigma$ &   5.27 days \\
Basic reproduction number & $R_0$ &   2.64  \\
Case-fatality rate & $\mu$ & $2.5\%$ \\
\bottomrule
\end{tabular}
\end{center}
\caption{Inference results of COVID-19 virus parameters without any interventions. According to model's estimations it takes around 16 days to recover after being infected, 5 days for symptoms to develop after the exposure and mortality rate is equal to 2.5\%.}
\label{tbl:virus-stats} 
\end{table}

%% file: 5_policy_strength.tex
\section{Estimation of policy strength by the Change-Point Model}
\label{sec:policy}
This section introduces a change-point methodology to quantify the efficiency of the major interventions applied worldwide to mitigate the COVID-19 spread. First, we briefly introduce the mathematical model behind the change-point method. Second, we continue with a brief description of a corresponding probabilistic programming model that detects change-points in the course of the caseload after the country applied policies and calculates the magnitude of the change. Next, we elaborate on several countries' data to conclude the efficiencies of investigated policies and give a summary of our findings in the final subsection.

\subsection{Mathematical Model of the Change-point Method}

\begin{figure}[h]
\centering
    \begin{subfigure}[b]{0.5\linewidth}
    \captionsetup{width=.95\linewidth}
    \includegraphics[width=\textwidth]{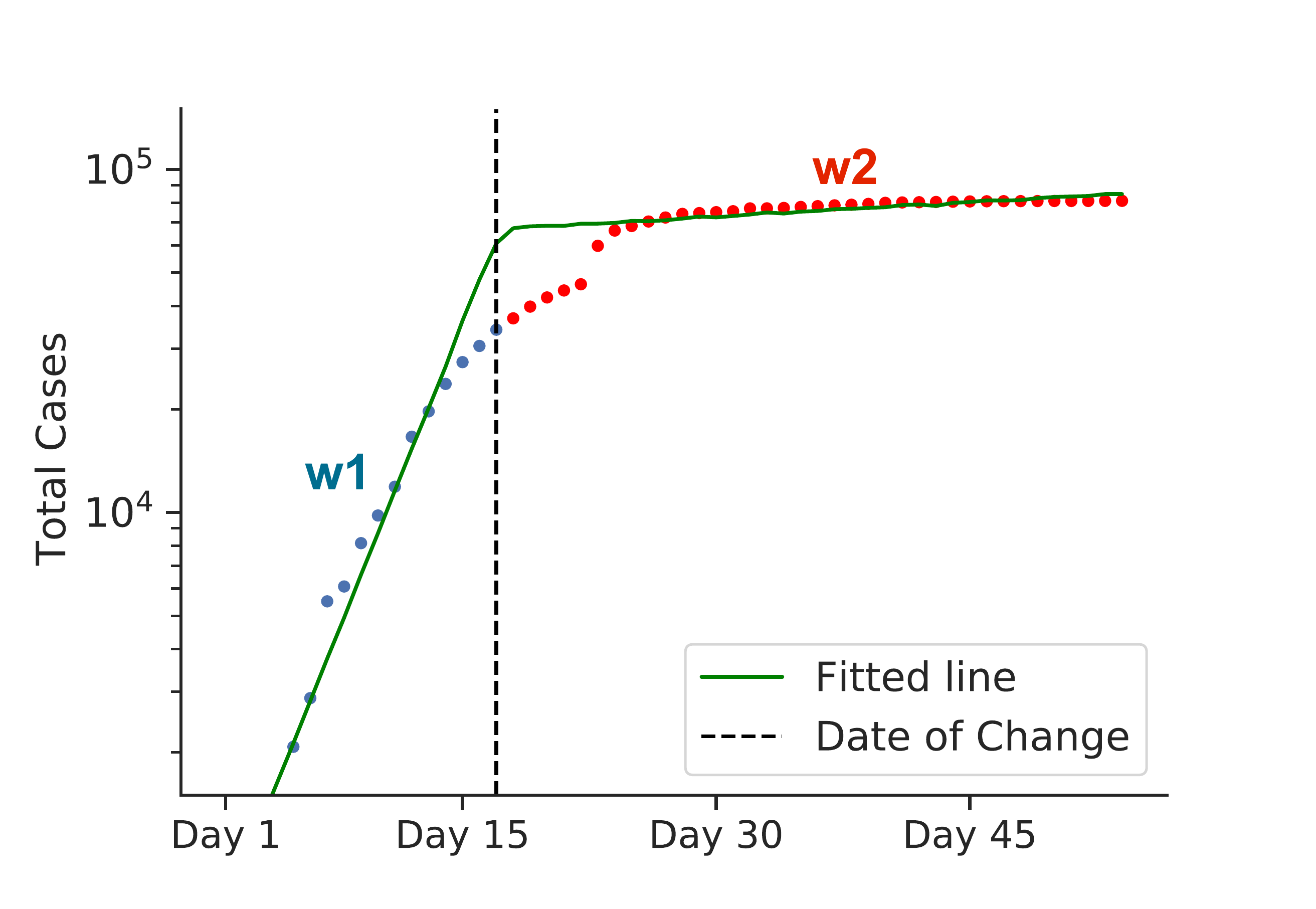}
    \end{subfigure}
    \caption{Graphical representation of a mathematical model. The green line represents a log-scaled case counts. Due to some interventions, the graph bends at the change-point, as indicated by the black dotted line. The change-point divides the graph into two intervals: blue and red. Both lines have different slopes, $w_1$ and $w_2$, respectively, which represent the virus transmission rates before and after the intervention took came into effect. To calculate the efficiency of the intervention, we use the formula $1-w_2/w_1$. The data for early Chinese cases was used for this example. Here, the calculated $1-w_2/w_1$ value represents the efficiency of the lockdown policy of China back in early 2020. \looseness=-1}
    \label{toy}
\end{figure}

The mathematical model of the change-point method is based on the SEIRD compartmental model (see \S~\ref{sub:seird}) with some simple assumptions and new tems we describe below. The transmission rate $\beta$ is the number of susceptible individuals that an infected individual can infect in a day, which is calculated as $\beta = R_e\gamma$. At the very beginning, almost everyone in our setting is in the Susceptible compartment, so we can assume that $S$ is equal to the total population size $N$, or $S=N$. With this approximation, Eq.~(\ref{eq:SEIR_S}) can be rewritten as follows:
\begin{align}
\frac{dS}{dt} &=-R_e \gamma I = -\beta I \label{eq:SEIR_S_approx}
\end{align}

Additionally, since the incubation period is much shorter than the recovery time, we can ignore the E, R, and D compartments at the initial stage of the simulation. Thus, we can approximate the total population size as $N = S+I$ and  Eq.~(\ref{eq:SEIR_I}) becomes:
\begin{align}
\frac{dI}{dt} &= \beta I \label{eq:SEIR_I_approx}
\end{align}

The size of the Infected compartment rises exponentially with the rate $w = \beta$ (that is, on day $t$, the number of infected cases is calculated as $e^{ \beta t}$). Due to the exponential nature, it is appropriate to investigate the case counts using a log scale. In the log scale, the exponential spread is represented as a linear line, with the transmission rate of the virus $\beta$ represented by the slope $w$.

Consider a toy example in Fig.~\ref{toy}.  In the beginning, the graph is a steep line with the slope $w_1$ (representing a rapid, exponential spread). After a corresponding policy is applied, the graph bends and becomes less steep with the slope $w_2 \ll w_1$ (slower spread). Therefore, the graph roughly consists of 2 lines of different slopes, $w_1$ and $w_2$, with a separation point in-between, which we call a \textit{change-point} (the black dotted vertical line in the graph). The slope $w_1$ and $w_2$ represent the transmission rates before and after the change-point. Since $w_1$ and $w_2$ represent the transmission rates before and after the policy takes its effect, we can define the strength of the target intervention in terms of their ratio\cite{ramkissoon_2020}:
\begin{equation}
\text{Policy efficiency} = 1 - \dfrac{\beta_2}{\beta_1} = 1-\dfrac{w_2}{w_1}
\label{eq:policy_strength}
\end{equation}



Given the incubation period, we expect the policy will show effect after around 2 to 4 weeks after the policy establishment. 

\subsection{Probabilistic Programming Model of the Change-point Method}
As was mentioned in the Related Work section (see \S~\ref{sec:related:probabilistic-algorithms}), probabilistic models were successfully used to detect change-points in transmission rates of coronavirus. In the present subsection, we describe the change-point model adapted from \cite{ramkissoon_2020}, which we used to compare the efficiencies of major policies applied by different countries.

\subsubsection{Likelihood Choice}
In our probabilistic setting, the likelihood corresponds to the log-scaled line of accumulated confirmed cases. Following the example of \cite{ramkissoon_2020}, we chose piece-wise linear regression and added the StudentT noise, which is more robust w.r.t the outliers than conventional Gaussian noise. We define $\tau$ as the change-point in the range $[0, 1]$, with 0 and 1 being the start and end of the simulation time period, respectively. The likelihood can be modeled as follows: 
\begin{equation}
y = wt + b + \epsilon,
\end{equation}
where 
\[
    w, b = 
\begin{cases}
    w_1, b_1 ,& \text{if } t < \tau\\
    w_2, b_2, & \text{if } t > \tau
\end{cases}
\]
\[
    \epsilon \sim StudentT(2,0, \sigma^2)
\]

Note that the weights $w_1$ and $w_2$ correspond to slopes before and after the change-point in Eq.(\ref{eq:policy_strength}) of our mathematical model. 
\newline

\subsubsection{Prior Choice}
Here we illustrate the choice of parameters' priors used as input for our probabilistic model to draw samples from, following the work \cite{ramkissoon_2020}. For weights, we use normal distribution, with $w_2$ having the mean equal to zero as we expect the slope to drop significantly after the change-point.
\begin{equation}
\begin{array}{l}
w_1 \sim N(0.5, 0.25) \\
w_2 \sim N(0, 0.25)
\end{array}
\end{equation}

For bias terms, we set the priors to be from normal distribution. However, this time, we adjust the bias priors for each country adaptively since bias is sensitive to each country's course of the caseload. Following \cite{ramkissoon_2020}, we assign the mean of y in the first and fourth quartiles to $m_1$ and $m_2$, respectively. For $b_1$ to be relatively flat the standard deviation $s_1$ is set to 1. $s_2$ is set to $0.25m_2$

\begin{equation}
\begin{array}{l}
b_1 \sim N(m_1, s_1) \\
b_2 \sim N(m_2, s_2)
\end{array}
\end{equation}

We use Beta distribution to model the change-point $\tau$, and assume that the change is more likely to occur in the second half of the date range.

\begin{equation}
\begin{array}{l}
\tau \sim Beta(4, 3)
\end{array}
\end{equation}




Lastly, we use the HMC algorithm with NUTS \cite{nuts} for the inference step, described in \S~\ref{subsub:mcmc}.

\subsection{Inferring efficiencies of major policies with the Change-point Method}
We investigate major initial interventions applied by several countries to mitigate the virus spread. For more accurate results, we chose nine countries presented in Table~\ref{tbl:policy} that strongly imposed corresponding policies, assuming that they were applied to the fullest extend. In this experiment, we focused on five main policy categories:
\begin{itemize}
    \item \textit{Lockdown}: A lockdown is an intervention that forces people to stay where they are. It includes a gathering ban, closure of non-critical services, and strict transportation restrictions. People cannot freely enter or exit their designated areas, and economic activities are essentially suspended.
    \item \textit{Social distancing}: Social distancing includes interventions or measures intended to maintain a physical distance between people, including a gathering limit or closure of non-essential services. It can be considered as a partial or a soft lockdown.
    \item\textit{Contact tracing and social distancing}: Contact tracing is the policy that investigates the close contacts of infected cases and then tests and quarantines them. Investigating the countries with successful contact tracing campaigns revealed that they coupled the contact tracing intervention with social distancing (e.g., South Korea, Australia, and Vietnam). For this reason, instead of addressing contact tracing separately, we merged it with social distancing to be closer to the real-world scenario.
    \item \textit{Mask and hygiene mandate}: Almost every country imposed a mask mandate sometime in their COVID-19 timeline response. Since it is always coupled with other restrictions, separating the effect of mask mandate from other interventions is a challenging task. Because the change-point method cannot be applied in this case, we proposed a different approach to this issue, discussed in \S~\ref{subsubsec:hygiene}.
    \item \textit{Vaccine}: 
With the recent roll-out of vaccines worldwide, we also examine the effect of recent vaccination campaigns in Israel and the US. 
The effectiveness of the vaccines depends on several factors, including the time it takes to approve, manufacture, and deliver them to the population, as well as improvements, and the development of other vaccine variations, and the proportion of the population vaccinated. While there are many reports on the effectiveness of several vaccines in laboratory settings~\cite{kim2021looking, jones2021sputnik}, the early effects of the vaccinations on a large scale have yet to be studied in more detail. 
Assuming that vaccination campaigns can result in the same effects on virus mitigation as any other policy that the government may enact, we also analyze the effectiveness of the vaccination programs by applying the same change-point model.
\end{itemize}

\begin{figure}[h]
\centering
    \begin{subfigure}[b]{0.48\linewidth}
    \captionsetup{width=.95\linewidth}
    \includegraphics[width=\textwidth]{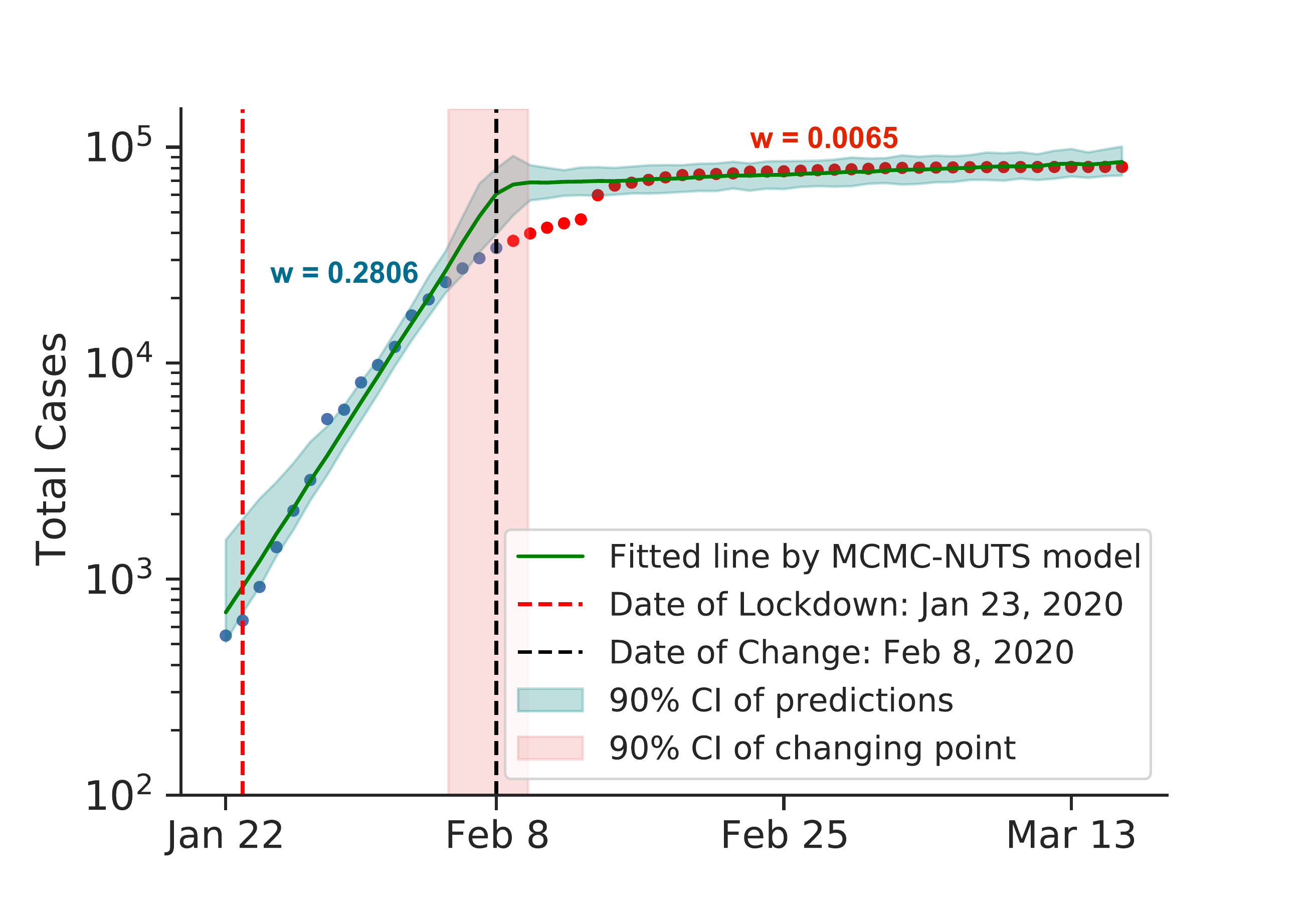}
    \caption{Fitted graph for China with change-point on February 8, 2020.}
    \end{subfigure}
    \begin{subfigure}[b]{0.48\linewidth}
    \captionsetup{width=.95\linewidth}
    \includegraphics[width=\textwidth]{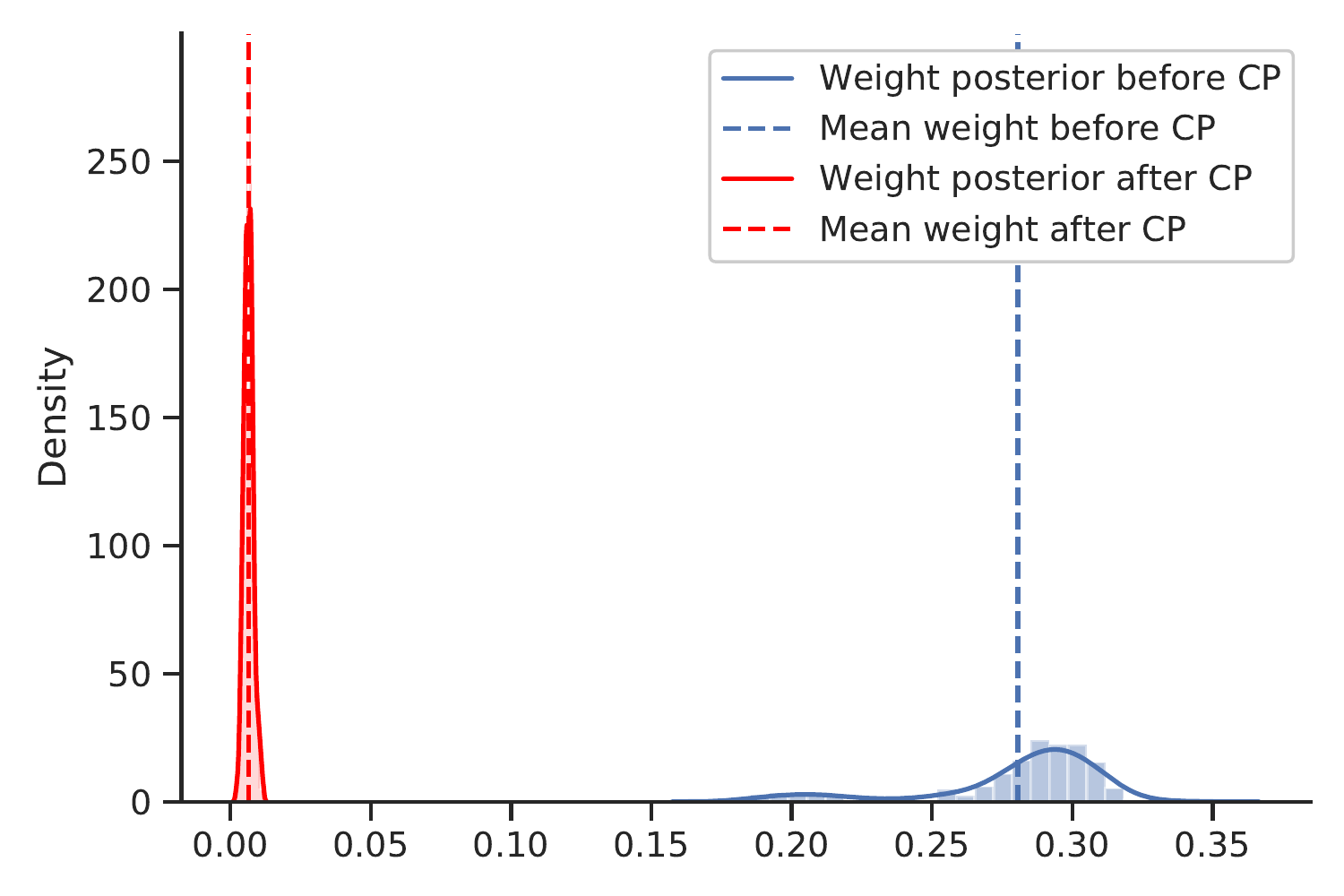}
    \caption{Posterior distribution of weights (transmission rate) before and after change-point.}
    \end{subfigure}
    \caption{Posterior for China. As the initial epicenter of the outbreak, cases in China skyrocketed in January 2020. China started applying a swift and stringent lockdown from January 23, 2020~\cite{wikipedia_china}, starting with the city of Wuhan, and managed to largely bring the outbreak under control in February, 2020. The growth rate was suppressed staggeringly from 0.2806 to a mere 0.0065, with the change-point estimated around February 8, 2020. \looseness=-1}
    \label{cn}
\end{figure}

\begin{figure}[h]
    \begin{subfigure}[b]{0.5\linewidth}
    \includegraphics[width=\textwidth]{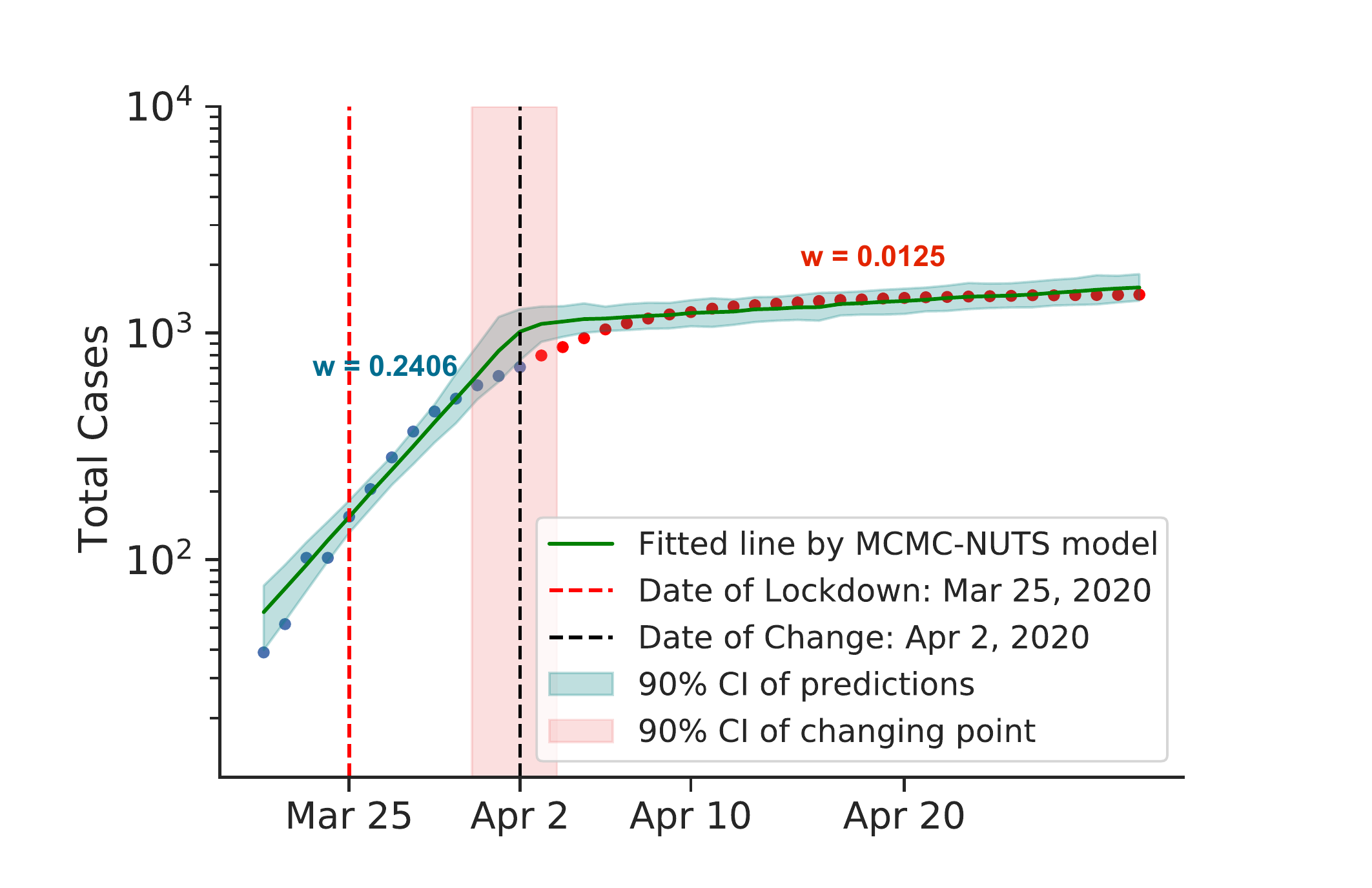}
    \captionsetup{width=.95\linewidth}
    \caption{Fitted graph for New Zealand with change-point on April 2, 2020. }
    \end{subfigure}
    \begin{subfigure}[b]{0.48\linewidth}
    \includegraphics[width=\textwidth]{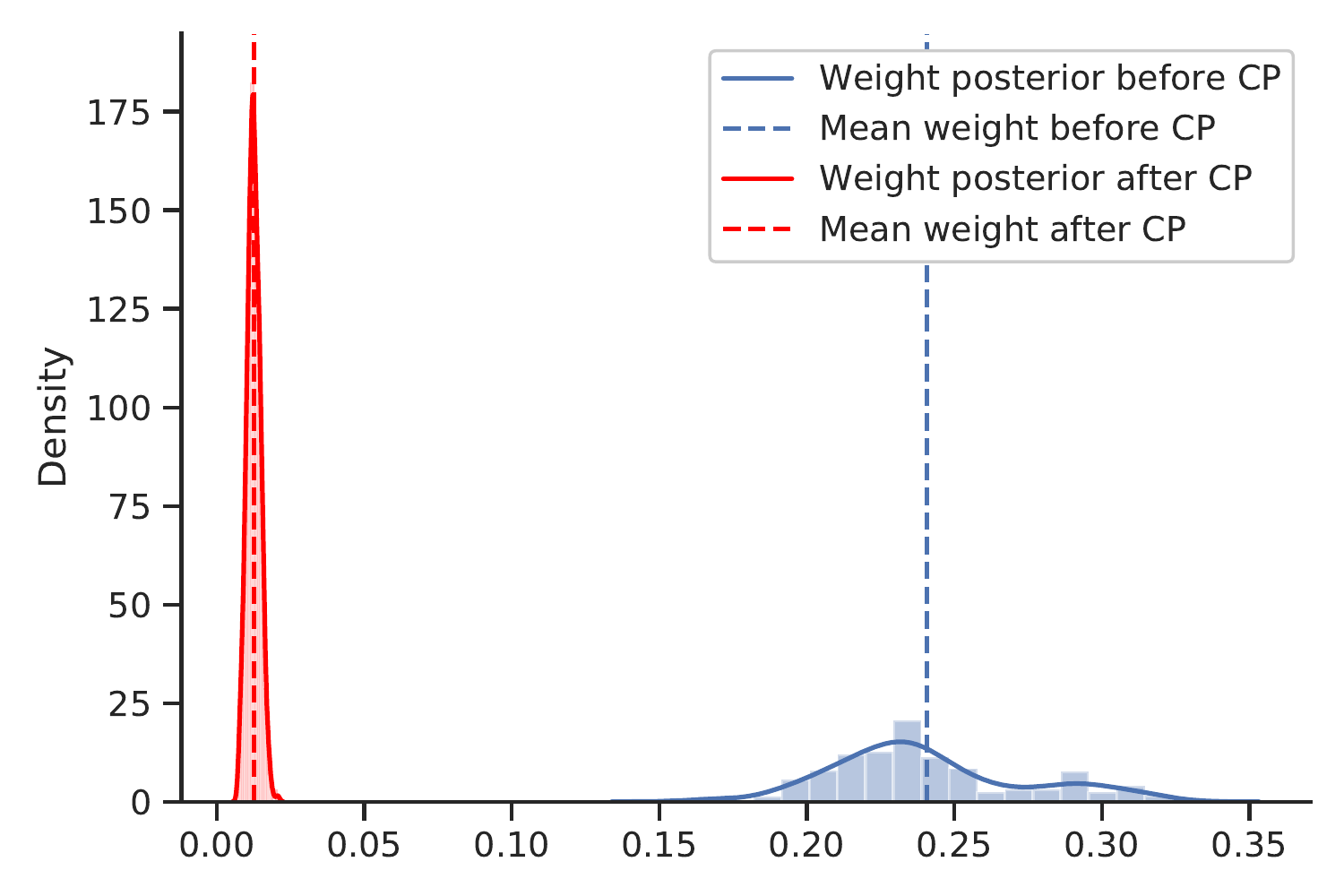}
    \captionsetup{width=.95\linewidth}
    \caption{Posterior distribution of weights (transmission rate) before and after change-point.}
    \end{subfigure}
    \caption{Posterior for New Zealand. After the confirming the first case, New Zealand experienced a rapid spread of the diseases. The government imposed a strong lockdown from March 25, 2020 \cite{wikipedia_new_zealand} and brought the outbreak under control in February, 2020. The growth rate declined from 0.2406 to 0.0125. The change-point was around April 2, 2020.}
    \label{nz}
\end{figure}

\begin{figure}[t]
    \begin{subfigure}[b]{0.48\linewidth}
    \captionsetup{width=.95\linewidth}
    \includegraphics[width=\textwidth]{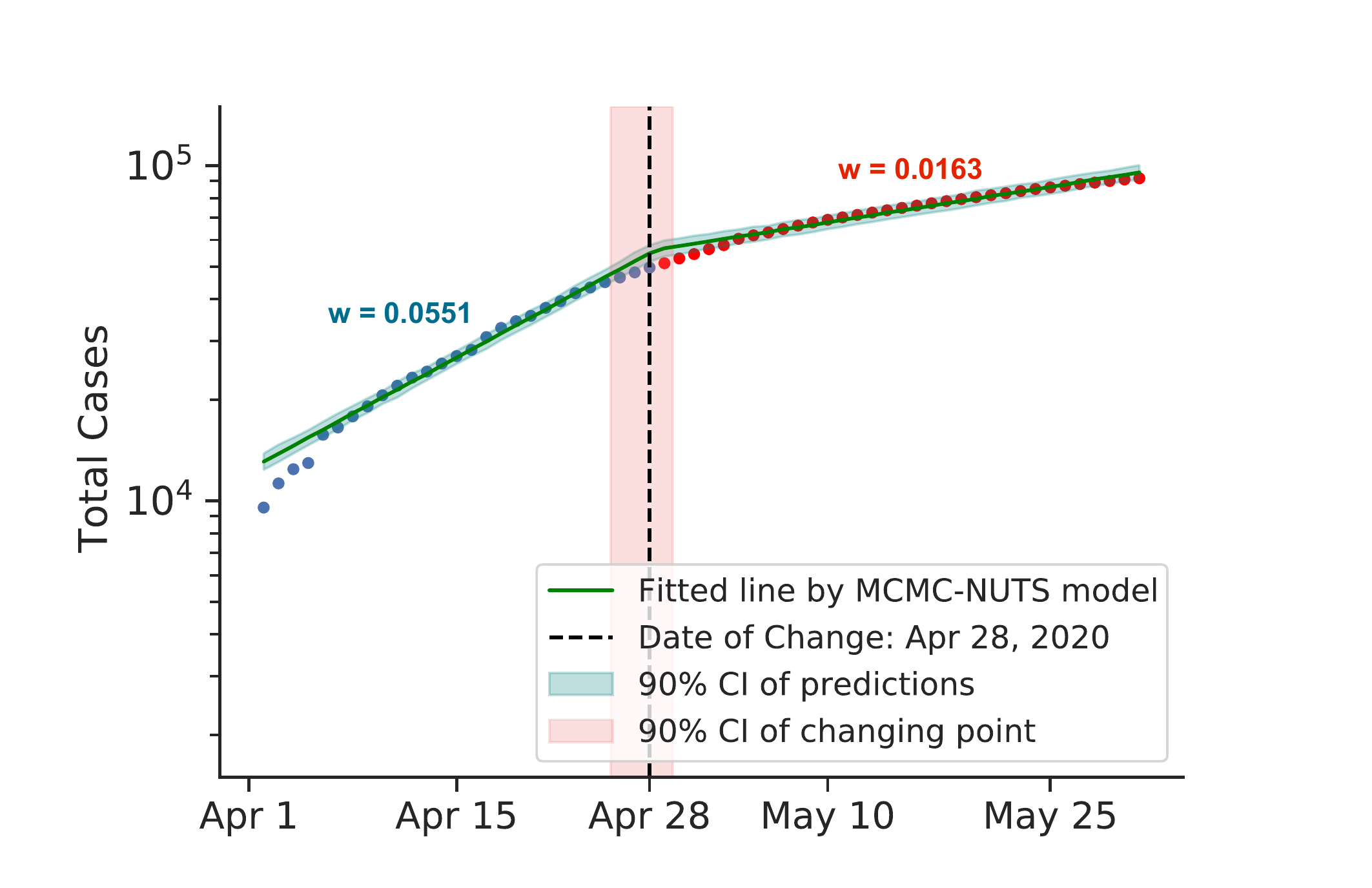}
    \caption{Fitted graph for Canada with change-point on April 28, 2020.}
    \end{subfigure}
    \begin{subfigure}[b]{0.48\linewidth}
    \captionsetup{width=.95\linewidth}
    \includegraphics[width=\textwidth]{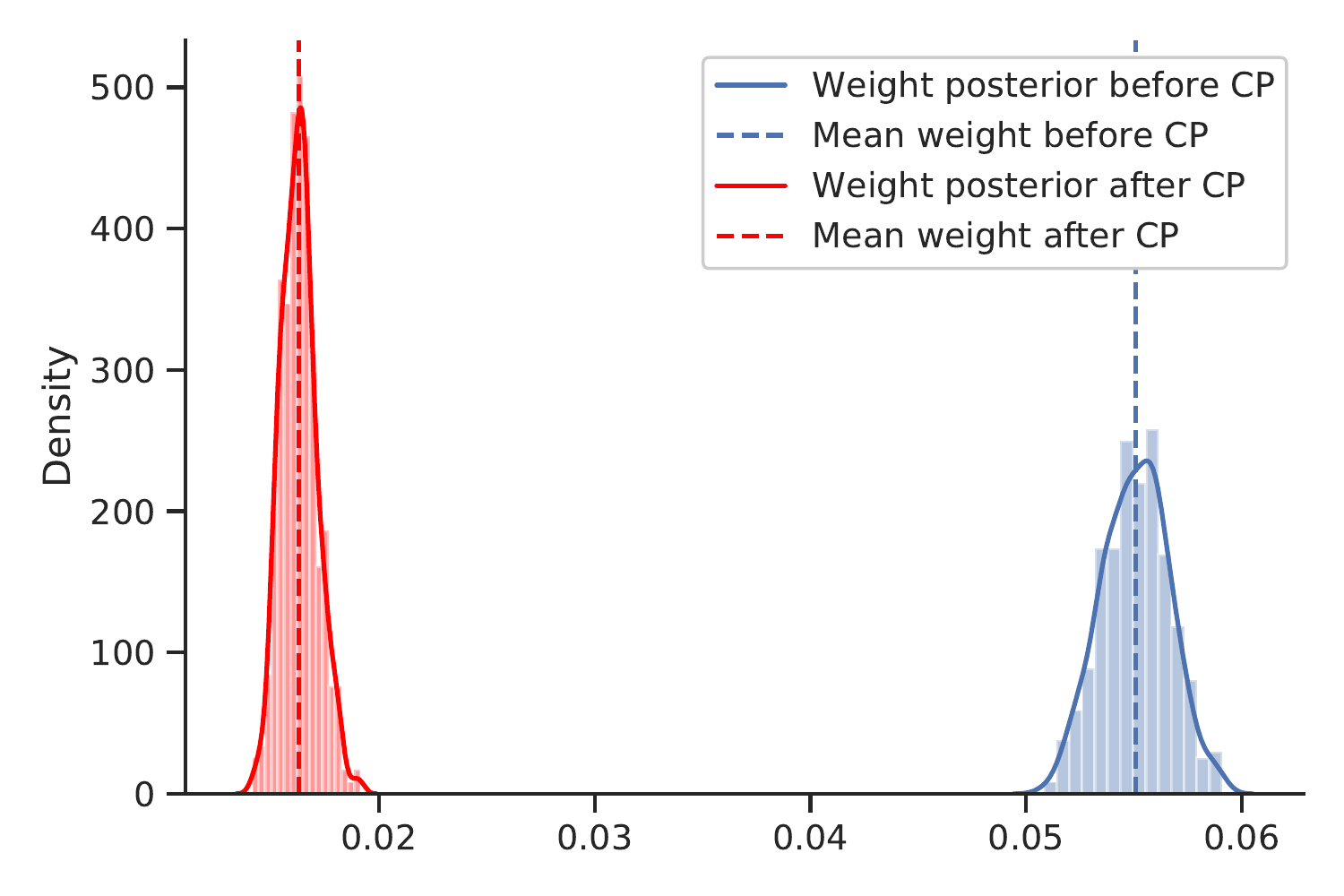}
    \caption{Posterior distribution of weights (transmission rate) before and after change-point.}
    \end{subfigure}
    \caption{Posterior for Canada. Canada recorded a total of more than 10,000 
    cases up to early April. The government imposed various restrictions in March-April \cite{wikipedia_canada} and reduced the transmission rate from 0.0551 to 0.0163. The change-point point was around April 28, 2020.}
    \label{ca}
\end{figure}

\begin{figure}[t]
    \begin{subfigure}[b]{0.48\linewidth}
    \captionsetup{width=.95\linewidth}
    \includegraphics[width=\textwidth]{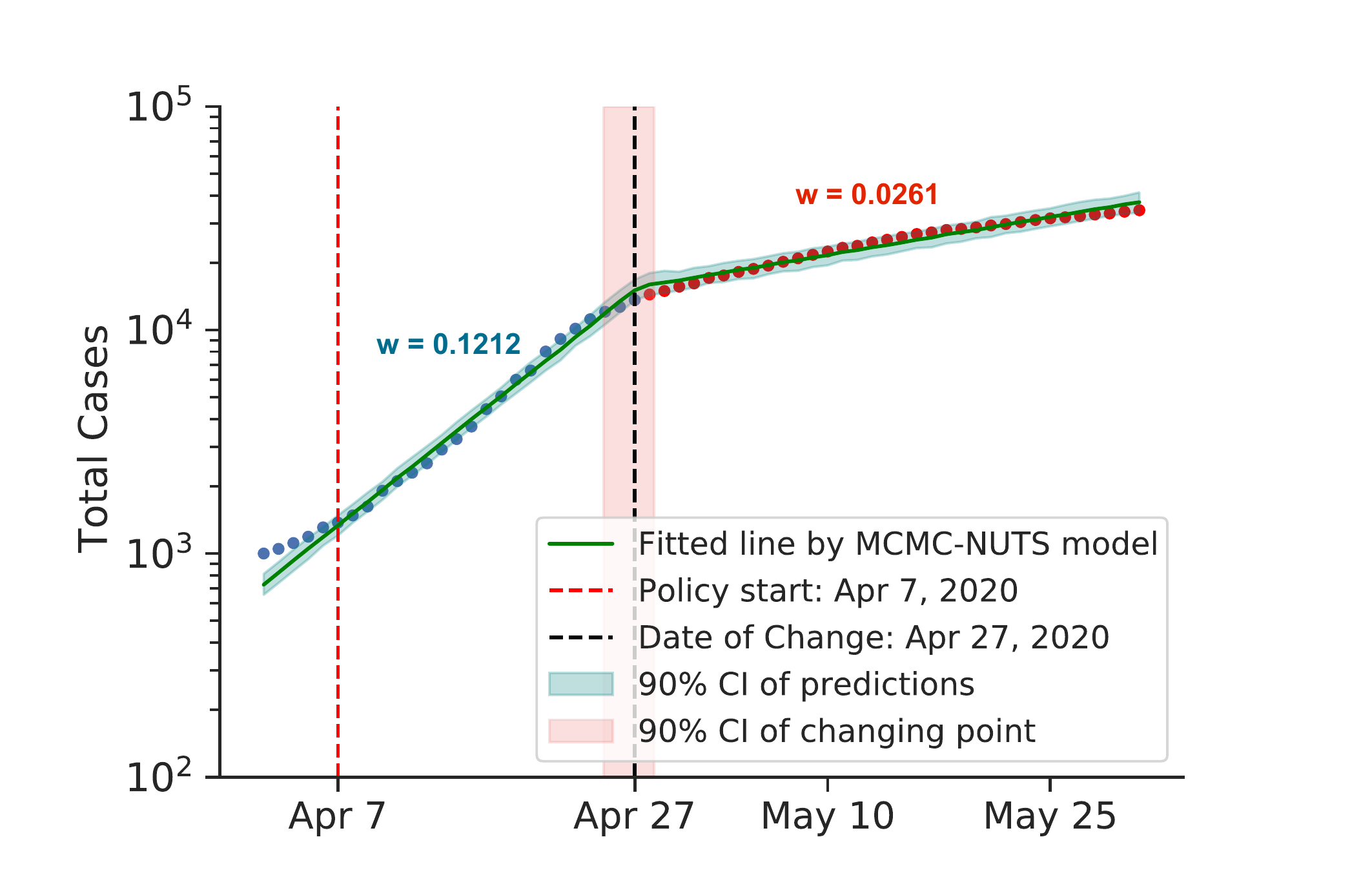}
    \caption{Fitted graph for Singapore with change-point on April 27, 2020.}
    \end{subfigure}
    \begin{subfigure}[b]{0.48\linewidth}
    \includegraphics[width=\textwidth]{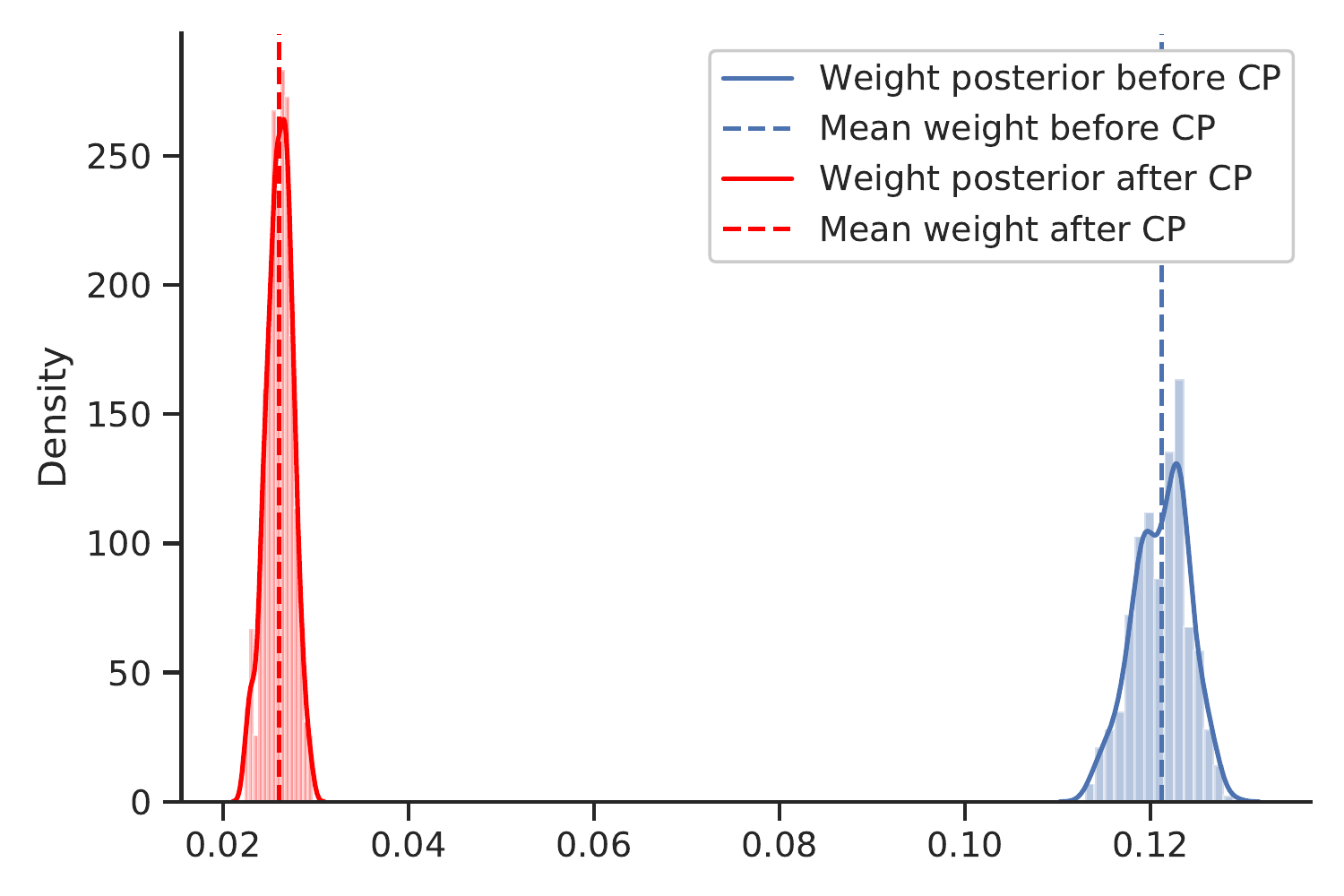}
    \captionsetup{width=.95\linewidth}
    \caption{Posterior distribution of weights (transmission rate) before and after change-point.}
    \end{subfigure}
    \caption{Posterior for Singapore. Singapore saw a rapid growth of the caseload among migrant workers around April 2020, and a 'circuit-breaker' was imposed on April 7 \cite{wikipedia_singapore} that brought the outbreak under control in late April, with the growth rate declining from 0.1212 to 0.0261. The change-point point was around April 27, 2020.}
    \label{sg}
\end{figure}

\begin{figure}[t]
    \begin{subfigure}[b]{0.48\linewidth}
    \includegraphics[width=\textwidth]{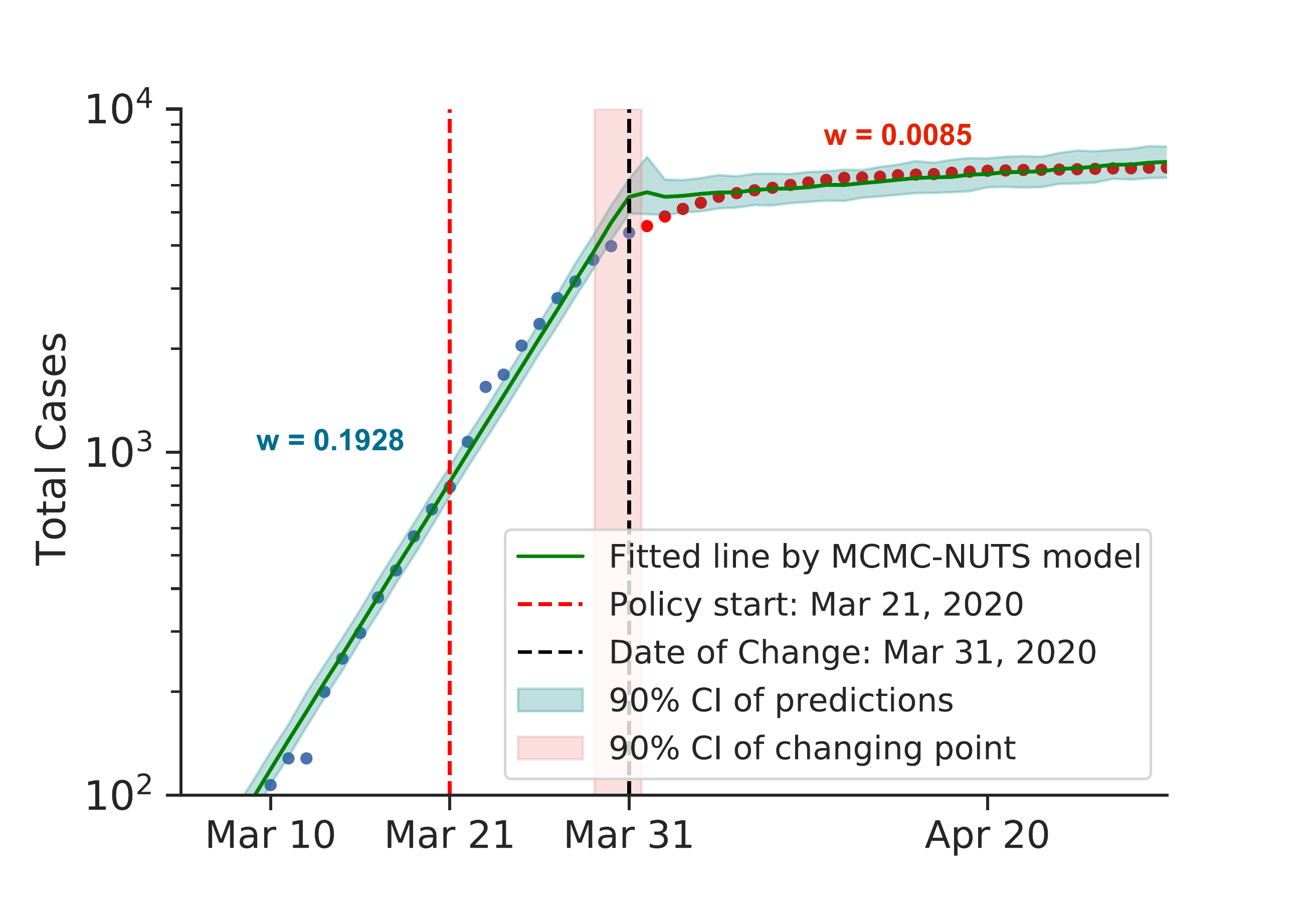}
    \captionsetup{width=.95\linewidth}
    \caption{Fitted graph for Australia with change-point on March 31.}
    \end{subfigure}
    \begin{subfigure}[b]{0.48\linewidth}
    \includegraphics[width=\textwidth]{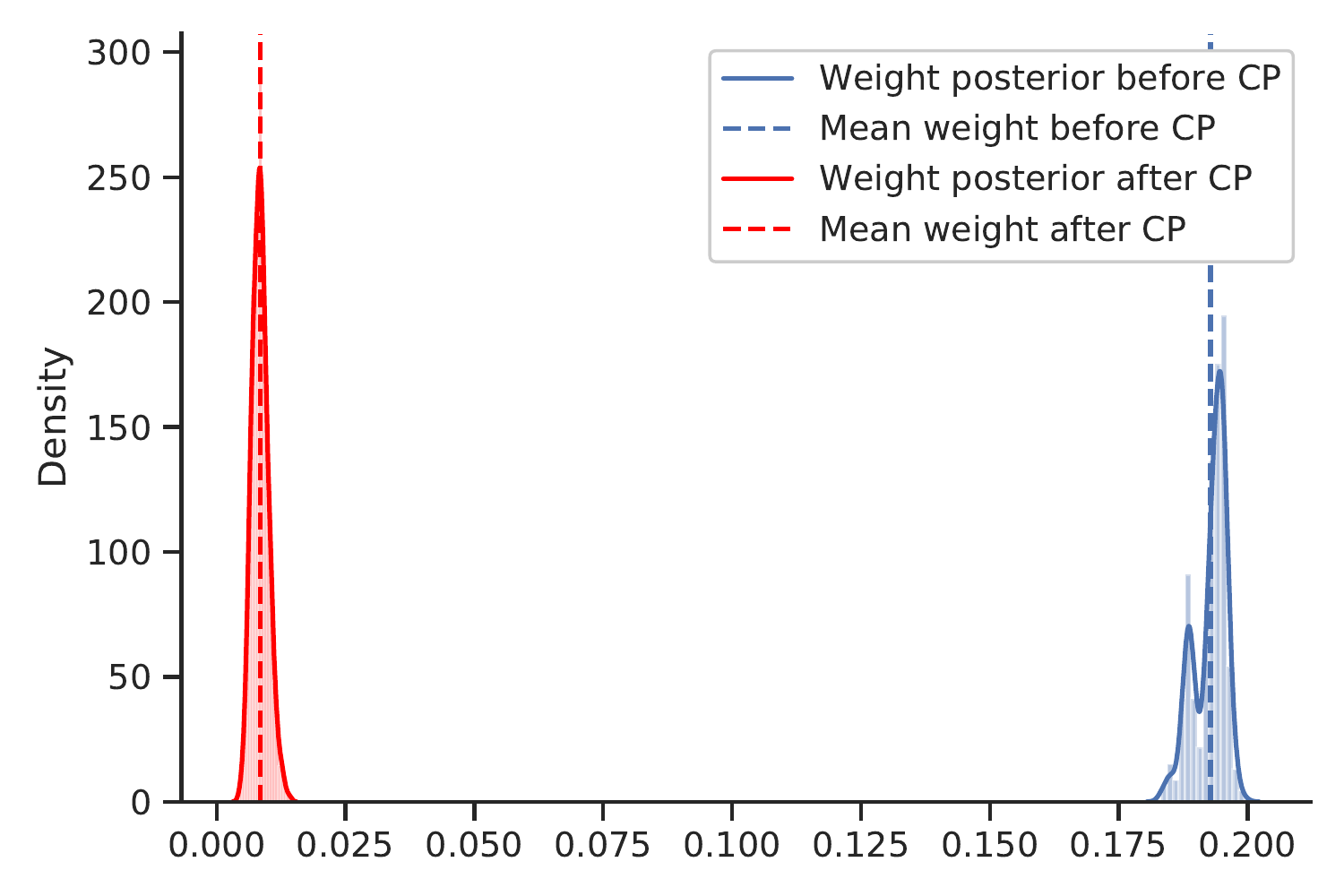}
    \captionsetup{width=.95\linewidth}
    \caption{Posterior distribution of weights (transmission rate) before and after change-point.}
    \end{subfigure}
 \caption{Posterior for Australia. After the confirming the first case, Australia recorded almost 10,000 cases before the end of March. With social distancing and contact tracing efforts from March 21, 2020 \cite{aus_covid}, Australia contained the spread, with a decline in the growth rate from 0.1928 to 0.0086. The change-point point was around March 31, 2020.}
    \label{au}
\end{figure}

\begin{figure}[t]
    \begin{subfigure}[b]{0.48\linewidth}
    \includegraphics[width=\textwidth]{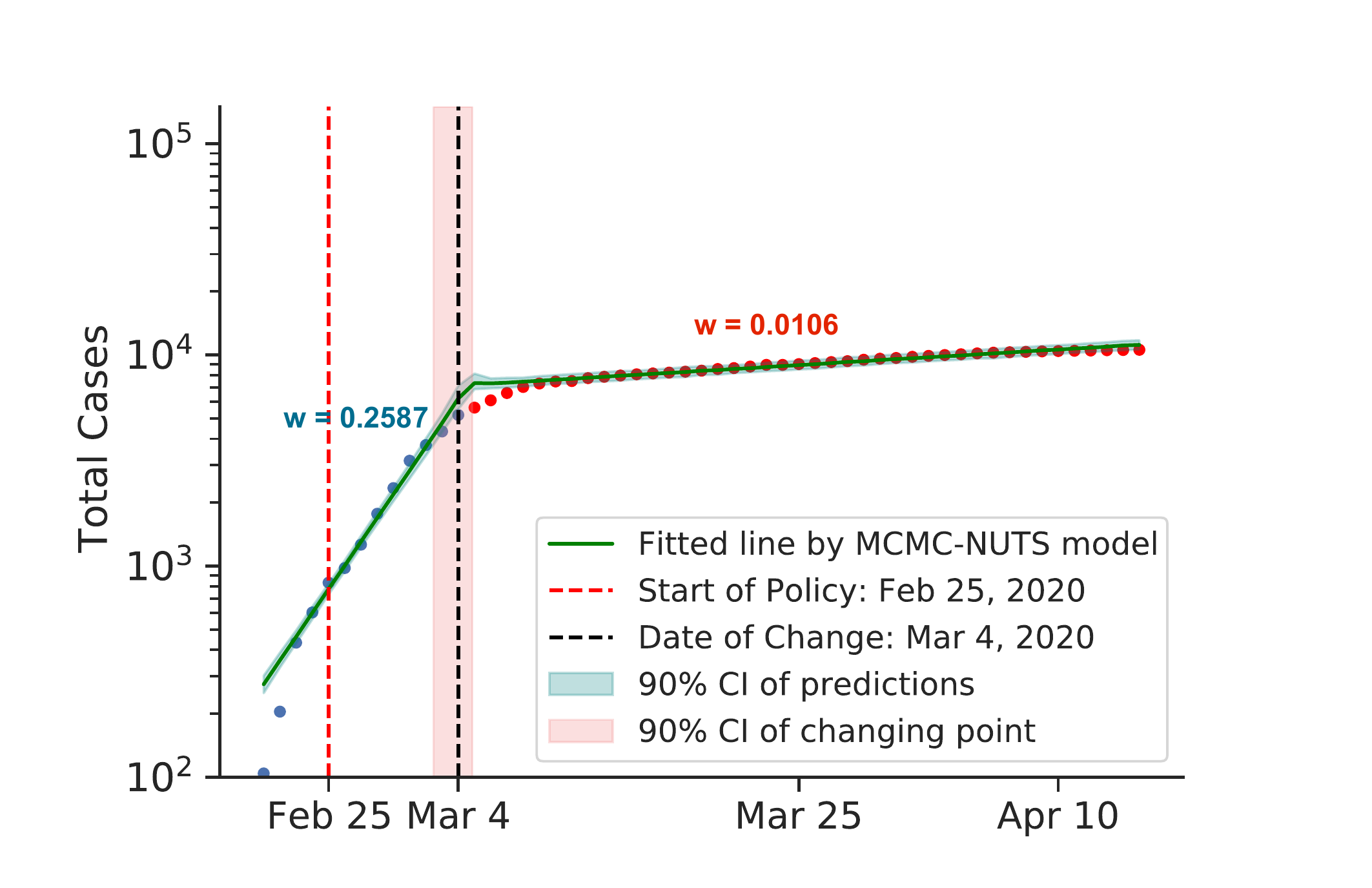}
    \captionsetup{width=.95\linewidth}
    \caption{Fitted graph for South Korea with change-point on March 4, 2020.}
    \end{subfigure}
    \begin{subfigure}[b]{0.48\linewidth}
    \includegraphics[width=\textwidth]{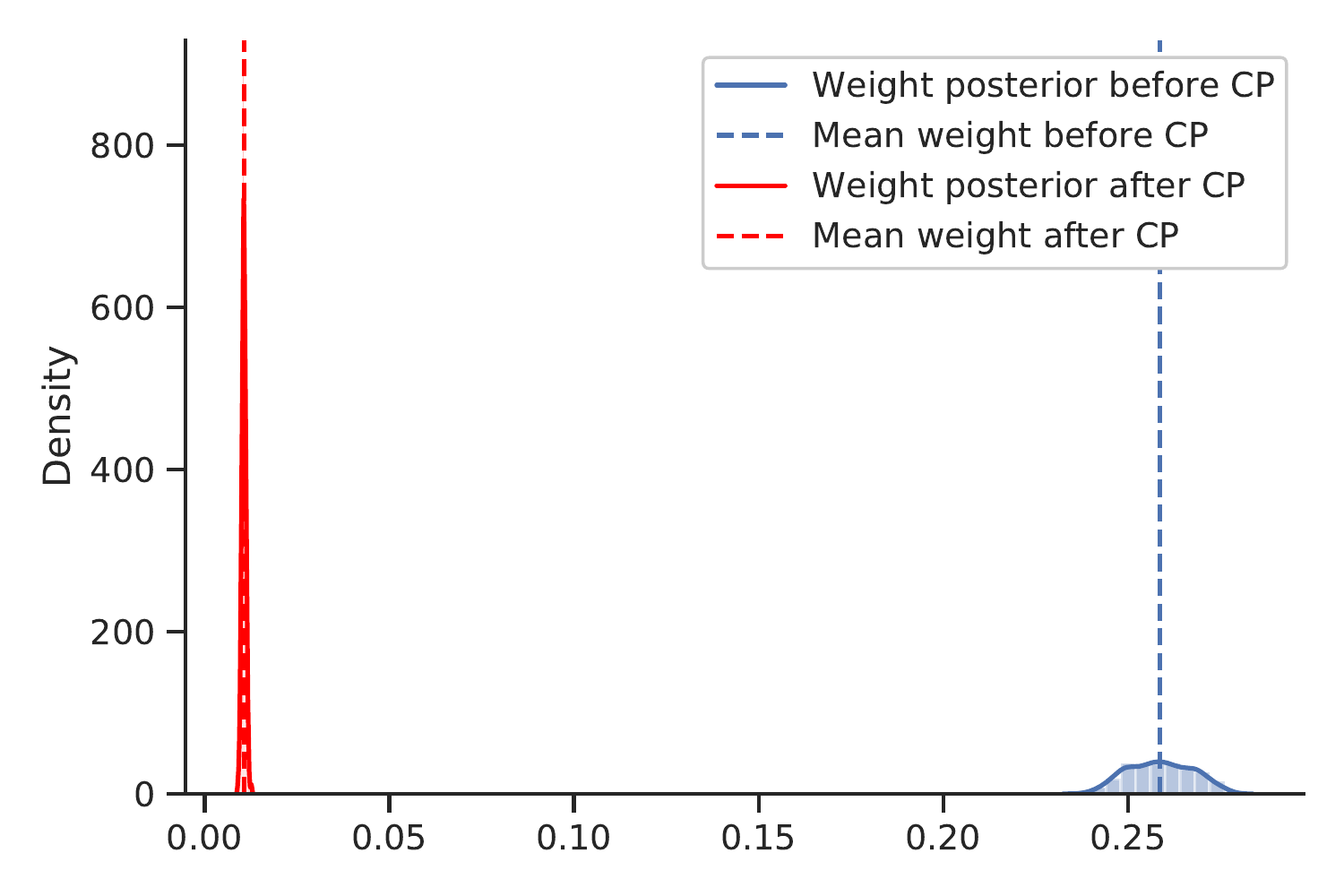}
    \captionsetup{width=.95\linewidth}
    \caption{Posterior distribution of weights (transmission rate) before and after change-point.}
    \end{subfigure}
    \caption{Posterior for South Korea. South Korea was the second epicenter of the outbreak after the super-spreader event of Patient 31. The government relied on social distancing and extensive contact tracing to avoid a stringent lockdown, starting shortly after the Patient 31 event (around February 25, 2020) \cite{kor_covid, lee2020testing}, and brought the outbreak under control in March, with a decline in the growth rate from 0.2587 to 0.0106. The change-point point was around March 4, 2020.}
    \label{kr}
\end{figure}

By using the probabilistic programming model described in the previous section, we detected the amount of time the policy needed to take effect after establishment, the change-point, and policy strength for each case. In all experiments, results have converged to the values consistent with our priors. The posteriors also fit well with the actual data (Fig.~\ref{cn}--\ref{kr}). The summary of policy efficiencies is shown in Table~\ref{tbl:policy}. In the next section, we discuss the results in more detail by investigating each policy separately by the country.


\begin{table}[ht]
\small
\centering
\begin{tabular}{l l r r r} 

\toprule
\textbf{Policy} &
\textbf{Country} &
\textbf{Started from} &
\textbf{Strength} &
\textbf{Take effect} \\
   \midrule
Lockdown & China & Jan 2020 & 0.98 & 16 days\\
Lockdown & New Zealand & Mar 2020 & 0.95 & 8 days\\
Contact tracing and distancing & South Korea & Feb 2020 & 0.96 & 8 days\\
Contact tracing and distancing & Australia & Mar 2020 & 0.96 & 10 days\\
Social distancing / soft lockdown & Canada & Mar 2020 &  0.70 & --\\
Social distancing / soft lockdown & Singapore & Mar 2020 &  0.78 & 20 days \\
Mask and hygiene & Japan & Feb 2020 &  0.30 & -- \\
Vaccine & US & Dec 2020 & 0.73 & 35 days \\
Vaccine & Israel & Dec 2020 & 0.88 & 59 days \\
No intervention (Herd immunity) & Sweden & Mar 2020 & 0 & -- \\
\bottomrule
\end{tabular}
\caption{Countries and policies used for analysis. Policy strength and time to take effect are inferred by change-point model. Countries and their major policies are listed according to the starting time and strength. In extreme policies such as lockdowns, the effect comes quickly, but there are side effects to be described later in \S~\ref{subsec:lockdown_fails}. }
\label{tbl:policy} 
\end{table}

\subsection{Discussion of the results of the Change-point Method}
\subsubsection{Lockdown}
We investigate the lockdown interventions imposed in China and New Zealand. Both countries applied a strict lockdown as their initial strategy to combat the virus spread.
The COVID-19 pandemic emerged in China, with the very first case confirmed on December 10, 2019 \cite{wikipedia_china}. New Zealand recorded its first case on February 28, 2020~\cite{wikipedia_new_zealand}. 


Both countries experienced a swift reduction in infection after the application of lockdown. With a strong centralized government, China could force a lockdown from January 23, 2020, starting with the epicenter of Wuhan city and Hubei province. The lockdown was overwhelmingly stringent, with a travel ban, a stay-at-home order, and transportation suspension. Other Chinese cities quickly followed suit with similar measures. Our model shows that the policy took its effect around February 8, 2020, with a 98\%  reduction in the transmission rate.

New Zealand recorded its first case on February 28, 2020. The New Zealand government introduced a four-tier alert level system and imposed a lockdown on most of the country's population and economy from March 25, 2020~\cite{wikipedia_new_zealand}. The policy seemed to take effect around April 2, 2020, with a 95\%  reduction in the infection rate. 

From these observations, we can conclude that a lockdown is capable of quickly curbing infections. We took the average efficacy of the two mentioned countries, 96\% as the efficacy of the lockdown for our further experiments.

\subsubsection{Social Distancing}
We investigated the social distancing imposed in Canada and Singapore. Both countries applied social distancing or soft lockdown mandates in their initial strategy to combat the virus spread.
The first COVID-19 case in Canada was confirmed on January 2, 2020 \cite{wikipedia_canada}. Around March--April, 2020, the Canadian government started to apply several restrictions to maintain social distancing \cite{wikipedia_canada}.

The first COVID-19 case in Singapore was confirmed on January 2, 2020 \cite{lee2020testing}. The government introduced a soft lockdown (dubbed a circuit-breaker), which included a stay-at-home order and cordon sanitaire\footnote{A cordon sanitaire is the restriction of movement of people into or out of a defined geographic area, such as a community, region, or country.}. Contact tracing was not extensively utilized until a later stage of the pandemic~\cite{wikipedia_singapore, wikipedia_singapore_main}. \looseness=-1

Both countries saw a considerable drop in the infection rate. Canada applied the restriction from around March to April, and the policy had an effect on around February 8, 2020, with a 70\% reduction in the infection rate. Singapore applied the circuit-breaker on April 7, 2020, and the change-point was determined to be on April 27, 2020, with a 78\% reduction in the infection rate.

It is evident that social distancing had a considerable effect on reducing the infection rate. We took the average efficacy of the two mentioned countries, 74\%, as the efficacy of social distancing.

\subsubsection{Contact Tracing and Social Distancing}
Australia and South Korea both utilized a contact tracing strategy coupling with social distancing as their initial strategy to combat the virus spread.

The first COVID-19 case in Australia was confirmed on January 25, 2020. On March 21, 2020, the Australian government imposed social distancing rules, with the closure of "non-essential" services. Swift recruitment of a large contact tracing workforce took place in March 2020 \cite{aus_covid}.

 In South Korea, the first COVID-19 case was confirmed on January 20, 2020~\cite{kor_covid}. The government raised the alert level to "Serious" on February 25, 2020, announced guidelines to limit trips and outdoor activities, and imposed emergency safety measures from basic hygiene rules to self-quarantine and social distancing \cite{lee2020testing}. Health officials implemented extensive movement and contact tracing to identify and inform exposed individuals \cite{lee2020testing}. 
 
Both countries experienced a swift reduction in infection after applying their social distancing coupling with contact tracing. In Australia, the policy seemed to take effect after 10 days (around March 31, 2020), with a 96\%  reduction in the infection rate. In South Korea, the policy showed effects on March 4, 2020, 8 days after the policy establishment, with an identical reduction in the infection rate.

It is evident that the social distancing when coupled with contact tracing can quickly curb the spread of infection. We take the average efficacy of the two mentioned countries, 96\%, as the overall efficacy. Thus, contact tracing could push the efficiency of social distancing to the same level as the lockdown.

\subsubsection{Effect of Mandating Masks and Hygiene} 
\label{subsubsec:hygiene}
We could not use the Changing-point model for masks and hygiene because they are hard to separate from other policies. However, we can indirectly represent the transmission rate via effective reproduction numbers. The transmission rate is proportional to reproduction number:
$$w = \gamma R_e$$

Therefore, we can use the reproduction number ratio.

$$\text{Policy efficiency} = 1 - \dfrac{\beta_2}{\beta_1} = 1 - \dfrac{w_2}{w_1} = 1 - \dfrac{R_{e_2}}{R_{e_1}}$$

We compared the effective reproduction number $R_e$ of Japan before policies were applied with the basic reproduction number $R_0$ of Sweden from Table~\ref{tbl:virus-stats}, which is equal to 2.64. 
 
The reason why we chose Japan lies in its cultural practices, which list the culture of wearing masks, very little physical contact (such as hugging or shaking hands), and not wearing shoes in the house \cite{japan_news}. We expect the reproduction number in Japan to be lower even if there is no strict policy applied. The average effective reproduction number $R_e$ for Japan after 6 runs was equal to 1.84. Thus, the efficiency of the hygiene and masks mandates is equal to $ 1 - 1.84/2.64 = 0.30$.

\subsubsection{Vaccine}
\label{subsec:vaccine} 
\begin{figure}[h]
\centering
    \begin{subfigure}[b]{0.48\linewidth}
    \captionsetup{width=.95\linewidth}
    \includegraphics[width=\textwidth]{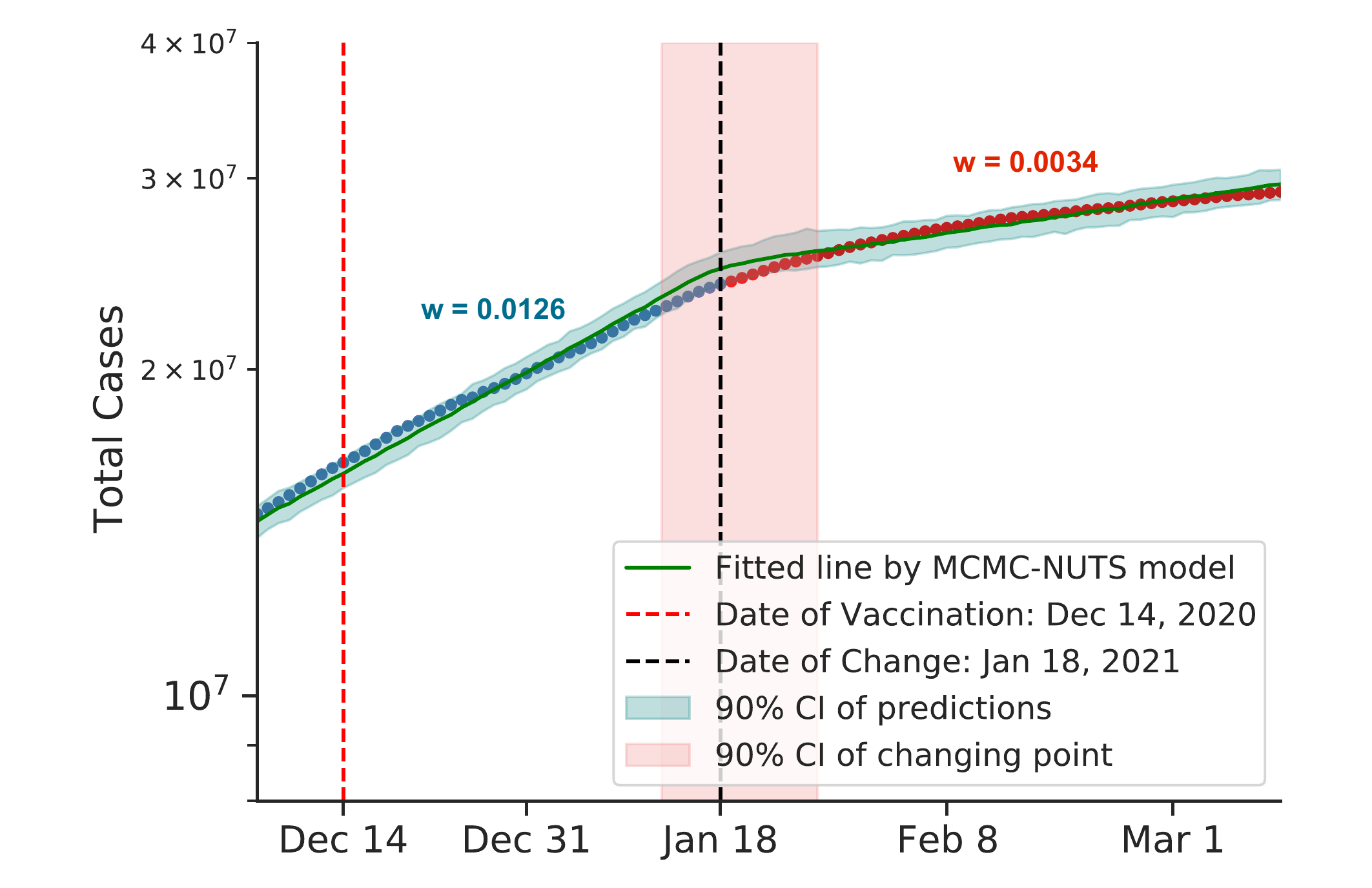}
    \caption{Fitted graph with change-point on January 18, 2021}
    \end{subfigure}
    \begin{subfigure}[b]{0.48\linewidth}
    \captionsetup{width=.95\linewidth}
    \includegraphics[width=\textwidth]{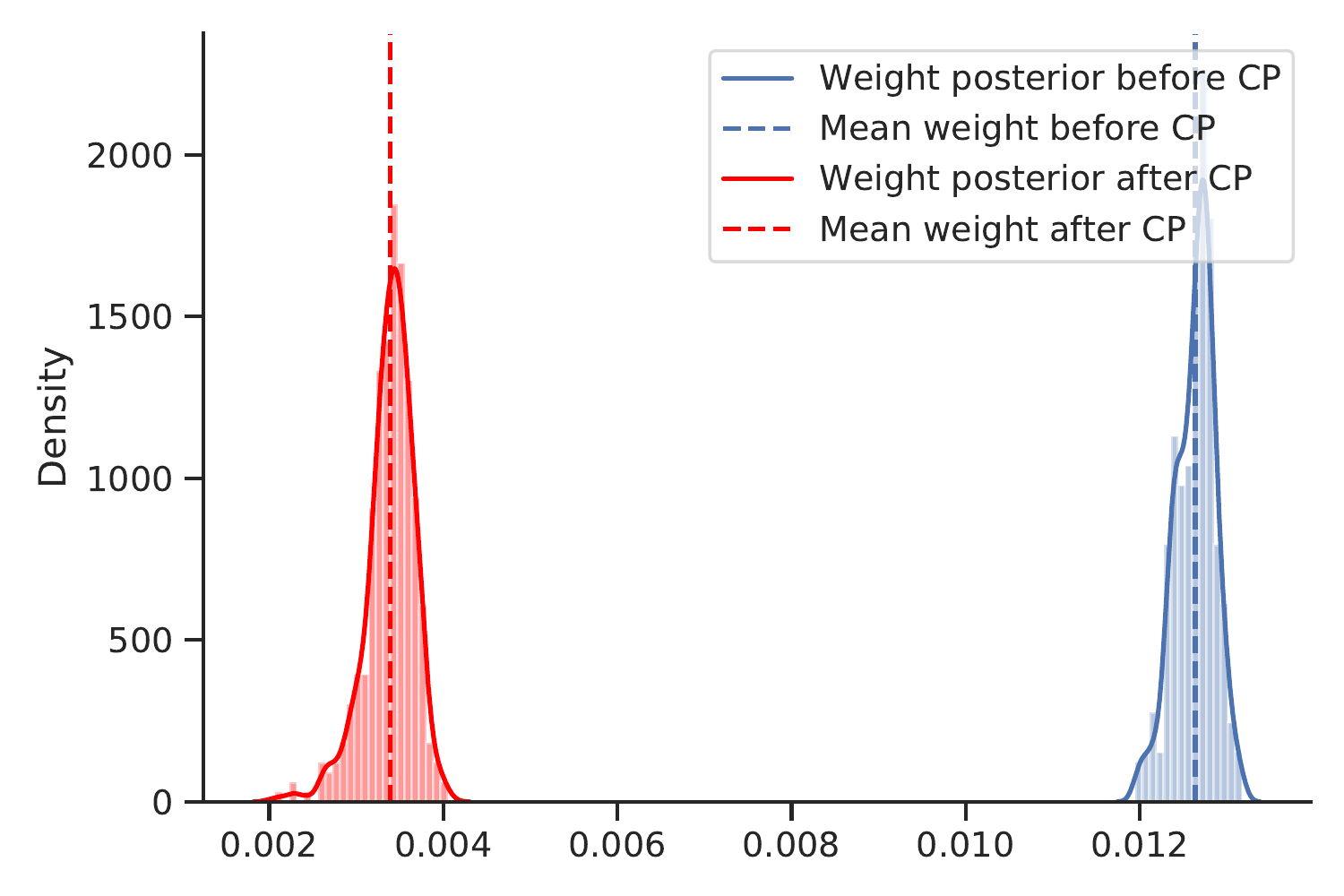}
    \caption{Posterior distribution of weights (transmission rate) before and after change-point.}
    \end{subfigure}
    \caption{Posterior for the US. The US recorded more than 20 million cases when they start vaccine program on December 14, 2020 \cite{painter2021demographic}, US sees the growth rate declined by 73\%, from 0.0126 to 0.0034. The change-point point was around January 18, 2021}
    \label{vaccine-us}
\end{figure}

\begin{figure}[h]
\centering
    \begin{subfigure}[b]{0.48\linewidth}
    \captionsetup{width=.95\linewidth}
    \includegraphics[width=\textwidth]{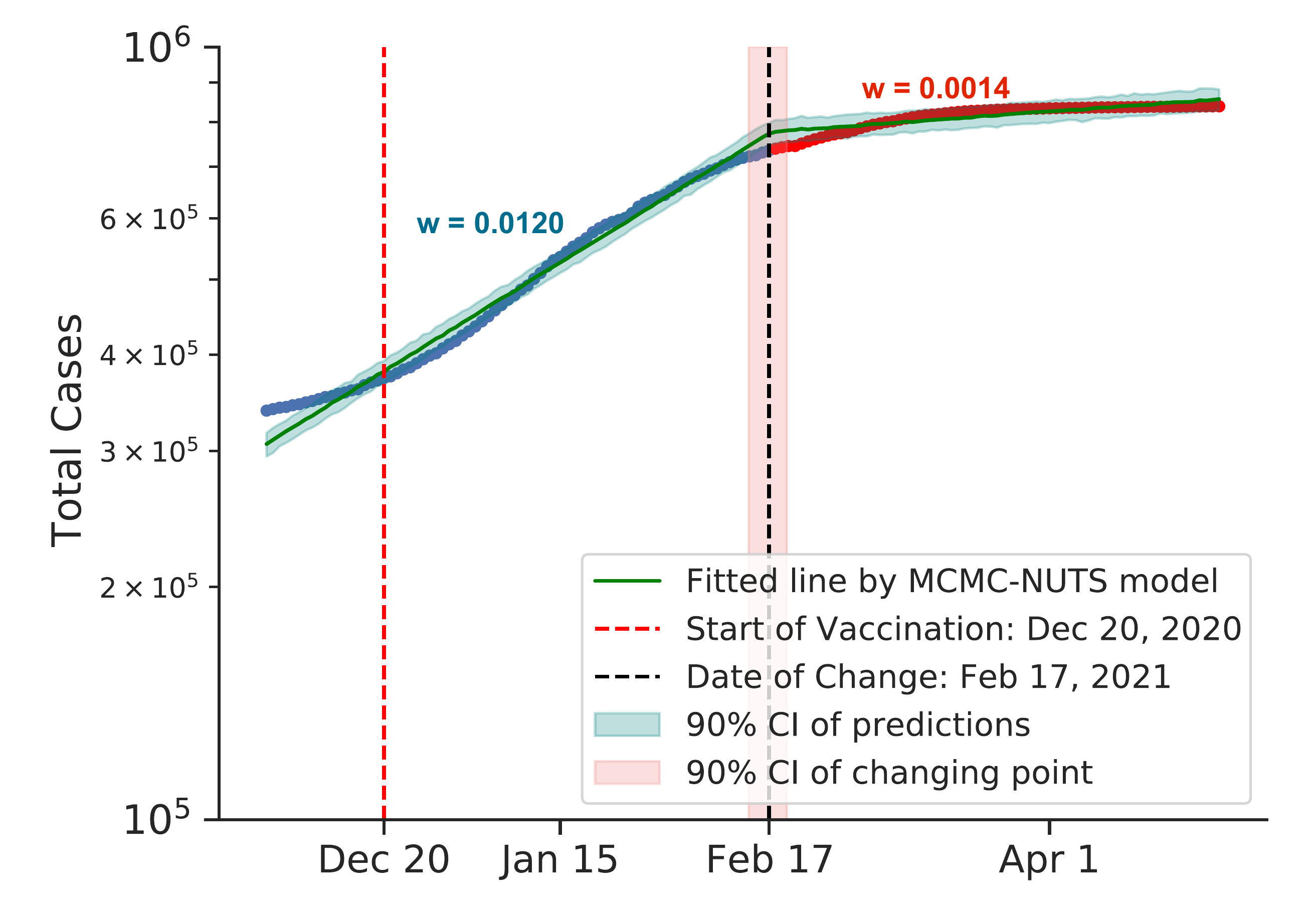}
    \caption{Fitted graph with change-point on February 17, 2021}
    \end{subfigure}
    \begin{subfigure}[b]{0.48\linewidth}
    \captionsetup{width=.95\linewidth}
    \includegraphics[width=\textwidth]{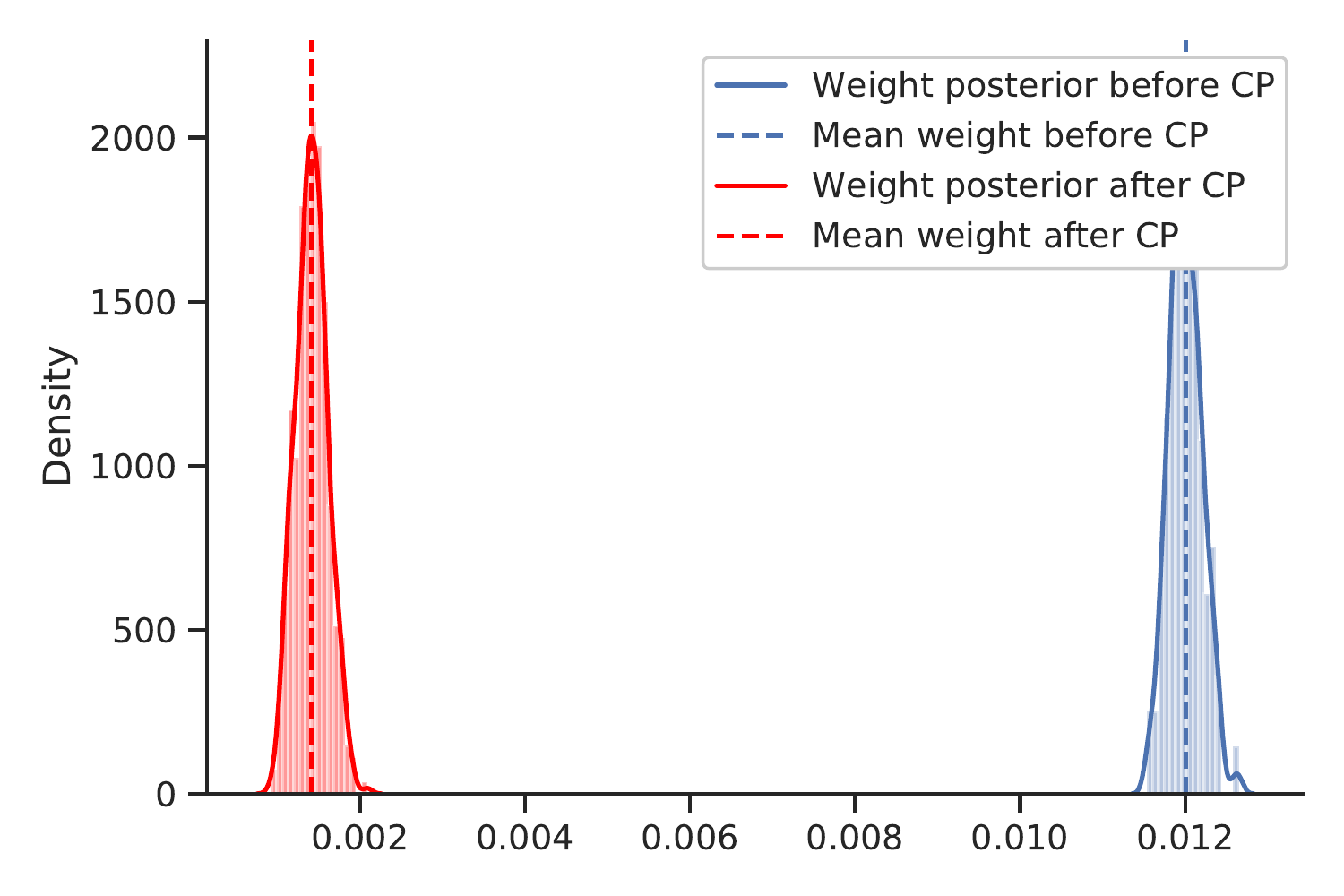}
    \caption{Posterior distribution of weights (transmission rate) before and after change-point.}
    \end{subfigure}
   \caption{Posterior for Israel. Israel recorded more than 30,000 cases when they started the vaccination program on December 20, 2020 \cite{rinott2021reduction}. After the vaccination took its affect, the growth rate declined by 88\%, from 0.0120 to 0.0014. The change-point point was around February 17, 2021.}
    \label{vaccine-israel}
\end{figure}

The US and Israel both have a sweeping and widespread vaccination program. The results obtained by our method for US and the Israel are plotted in Figures~\ref{vaccine-us} and \ref{vaccine-israel}, respectively. The US started the vaccine program on December 14, 2020 \cite{painter2021demographic}, while Israel started their campaign on December 20, 2020 \cite{rinott2021reduction}. These two countries experienced a swift reduction in infection after the vaccine program started. In the US, the policy seemed to show effect around January 18, 2021 with a 73\%  reduction in the infection rate. In Israel, the policy produced effects on February 18, 2021 with an 88\% reduction in transmission rate.  According to our results, in both cases, vaccination successfully mitigated the virus spread.

\begin{figure}[h]
\centering
    \begin{subfigure}[b]{0.48\linewidth}
    \includegraphics[width=\textwidth]{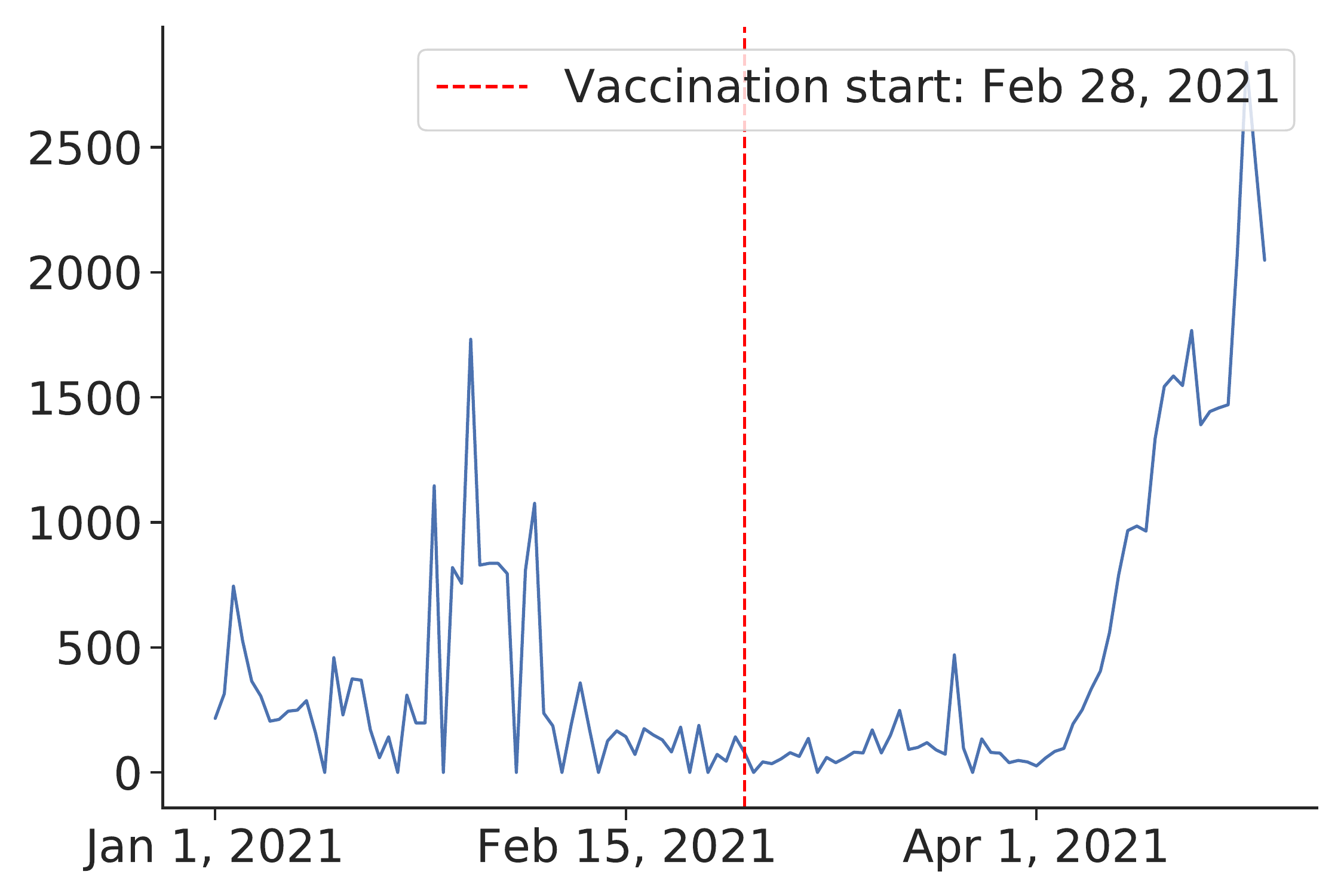}
    \caption{Thailand after vaccination on Feb 28.}
    \end{subfigure}
    \begin{subfigure}[b]{0.48\linewidth}
    \includegraphics[width=\textwidth]{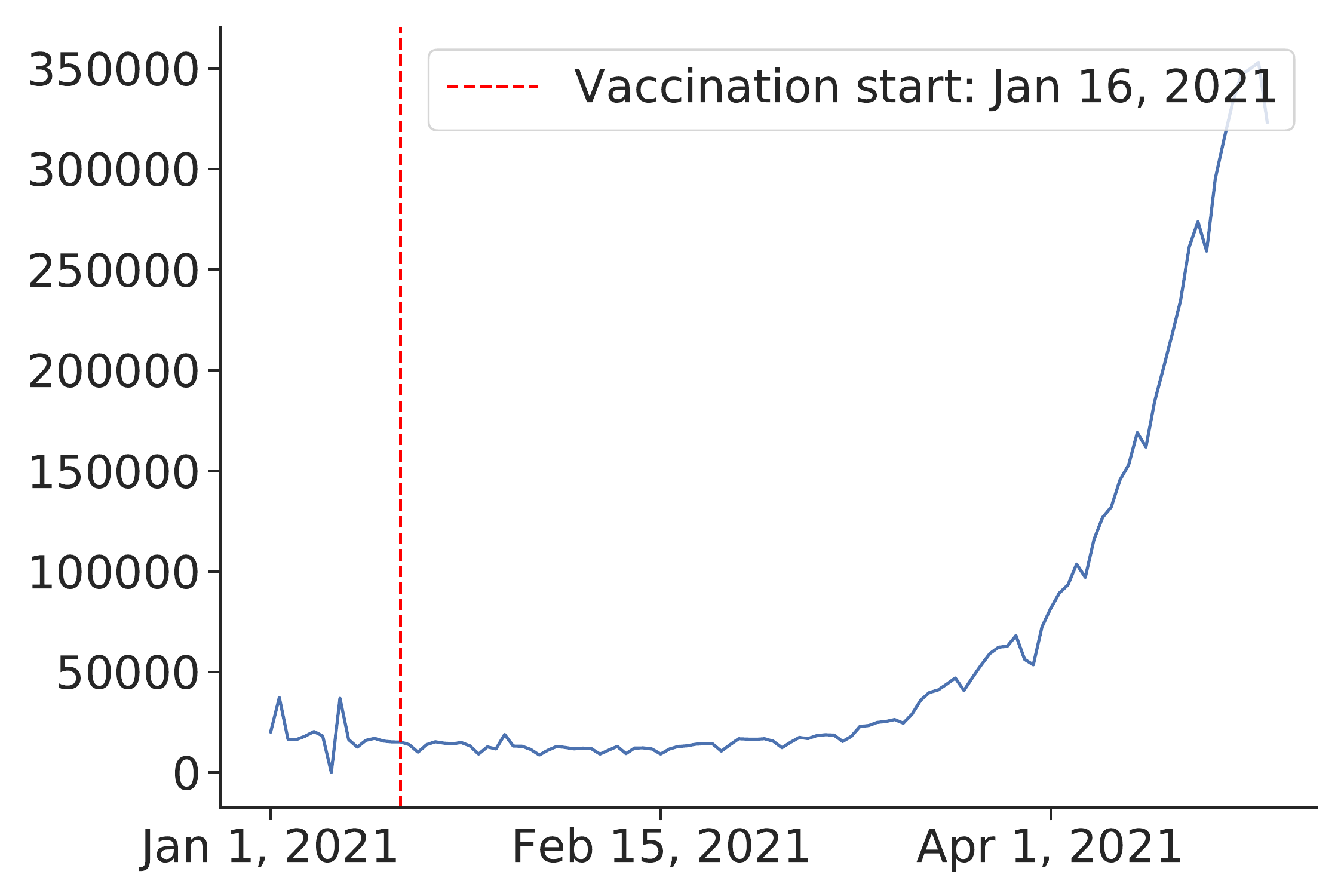}
    \caption{India after vaccination campaign on Jan 16.}
    \end{subfigure}
    \caption{Vaccine fails to mitigate virus in Thailand and India.}
    \label{vaccine-fail}
\end{figure}

However, most countries lack a swift and large-scale vaccination due to different reasons, including the delay in vaccine production, financial difficulties, or vaccine hesitancy \cite{WOUTERS20211023}. Thus, in most countries, the fraction of vaccinated population fall far below the herd immunity threshold according to the current data \footnote{\url{https://ourworldindata.org/grapher/share-people-fully-vaccinated-covid}}. The starts of vaccination programs can also leads to some incautiousness and fatigue that may have already driven up cases in many countries like India and Thailand. From Figure.~\ref{vaccine-fail}, we can see that after the vaccination campaign started ~\cite{thai_covid,india_covid} the number of cases increased drastically. It is possible that the reason for such an outcome lies in weakening awareness of coronavirus in the population after the vaccination campaigns start. People may have developed more relaxed attitude towards restrictions, which consequently may have caused these spikes in confirmed cases. We conclude that large-scale campaigns and accountability of the population in vaccination establishment play a key role in its success.

\subsection{Summary of Change-point Method's Results \& Limitations}
\subsubsection{Policy overview.}
In Section~\ref{sec:policy}, we evaluated the effectiveness of major policies based on the observed statistics. We found that social distancing, lockdown, and contact tracing are all effective in controlling the pandemic, with lockdown having the highest impact on the transmission rate (on average 96\% efficacy for China and New Zealand). It was also found that a combination of social distancing with contact tracing was shown to have an effect comparable to the lockdown  (also 96\% efficacy for South Korea and Singapore). The policy usually takes effect from 8-20 days after enforcement.

We also estimated the vaccination campaigns' efficiency. We found that although in countries like Israel and the US, vaccination effectively mitigated the virus spread (on average 81\% efficacy for Israel and the US), other countries like Thailand and India failed to bring virus spread under control. Moreover, it seems that vaccination programs were followed by a rapid increase in the confirmed case statistics in such countries. We suppose that reason for such controversial behavior lies in the lack of a large-scale vaccination program, as well as differences in public responsibility awareness.

\subsubsection{Limitations.}
This analysis was based on assumptions, where we boldly ignore the inherent differences between countries and populations. Complex factors such as the acceptance or awareness of the general public could affect the policy's effectiveness. It is evident that some Asian countries tend to perform better in containing the diseases, which we attribute to the collectivist nature and (usually) centralized government. For example, countries with experience with previous epidemics (China and Vietnam with SARS and South Korea with MERS) also tended to perform better thanks to previous experience in handling similar outbreaks. 

However, it is too early to conclude that stringent policies like lockdowns are the most successful at mitigating the COVID-19 pandemic, since the side effect of applying the policy should also be considered. Considering that the most efficient policies by our estimations may not be the most effective ones in terms of economic cost, we conducted additional experiments to address this issue. \looseness=-1

%% file: 6_simulation.tex
\section{Simulation by Generative Model}
\label{sec:simulation}

Having the virus statistics and policy capacities, we are ready to run our simulation experiments. Since all variables are already inferred, we can use a simple generative model to predict how pandemic plays out in different scenarios. To address the trade-off between public health protection and economic loss, we estimate the cost of the policies and the total loss for given caseloads and death tolls. We tried out multiple policy combinations to figure out what might be the best policy to fight the pandemic in our experimental setting.

\subsection{Model}

To simulate the infection and fatality cases, we used the SEIRD (see Eq.~\ref{eq:SEIR_D}). We follow the differential equations Eq.~\ref{eq:SEIR_S}--\ref{eq:SEIR_N} and the virus and policy statistic derived in \S~\ref{sec:compartmental}--\ref{sec:policy}. However, the fatality rate will not stay constant as we considered the hospital capacity.

\subsection{Assumption}
We used an imaginary country with a population of 1 million. In addition to the parameters we inferred from previous results, we applied some additional assumptions:
\begin{itemize}
    \item The hospital capacity is 60 per 100,000 capita (0.06\%). 
    Among OECD countries, the number of critical care beds ranges from 3.3 to 33.9 beds per 100,000 capita \cite{oecd}, and the number of hospital beds ranges from 50 to 1,300 beds per 100,000 capita \cite{oecd_instance}. Since countries might adapt normal bed into critical care beds to treat COVID-19 patients amid the health crisis, we use 60 critical care beds per 100,000 capita, which is already double the figure for the most resourceful country (33.9 for Germany)
    
    \item 6\% of the total cases are required to stay in the intensive care units (ICU). 
    
    Preliminary data on a subset of 7,162 COVID-19 patients age 19 and older with known health history in the US, from November 12, 2020 to March 28, 2020, found that 6\% requires ICU treatment.\footnote{\url{https://www.tfah.org/wp-content/uploads/2020/04/COVIDunderlyingconditions040320.pdf}} 
    
    \item ICU-required cases will die without ICU treatment. With treatment, the death rate for cases admitted to ICU is 60\%.
    
    Data from Washington, Seattle, and California suggests mortality rates reported in patients with severe COVID-19 in the ICU range from 50–65\% \cite{olicu}. 
\end{itemize}

We also estimate the economic and human capital cost for each policy:
\begin{itemize}
\item Lockdown: 10\% of GDP per year
\item Distancing: 5\% of GDP per year
\item Hygiene \& masks: \$2 per day per capita.
\item Infection: \$300 per infection per day (until recovered).
\item Contact-tracing: \$6,400 per new case.
\item Death: \$7 million per death
\end{itemize}

These estimations are reasonably set based on the following facts:
\begin{itemize}
\item Research suggests that global output shrinks by about 33\% at the peak of a lockdown, with an annual impact of over 9\% of the annual GDP \cite{mandel2020economic}.
\item The value of one human life is estimated to be A\$5.0M (\$3.48M) in Australia in 2020 \cite{au_cost} and \$10M in America in 2017 \cite{kniesner2019value}.
\item In South Korea, the treatment’s average daily cost for a mild patient is 180,000--260,000₩ (\$158--\$229) and for severe patients is 650,000₩ (\$572) \cite{han-soo_2020}.

\item  The costs of a contact tracing policy include the administrative (monetary) cost and the total quarantine days of the second-generation contacts. The standard contact tracing policy, where all close contacts are requested to quarantine for 14 days from the day of exposure, is estimated to cost 62.1 quarantine days and \$189 per index case~\cite{zimmermann}. We assume contact tracing is done on every confirmed case, and each quarantine day costs \$100 (South Korea’s government quarantine facility costs 100,000--150,000₩ (\$88--\$130) per day).\footnote{\url{https://kr.usembassy.gov/022420-covid-19-information/}} The total costs can be estimated as about \$6,400 per case (\$6,210 for 62.1 quarantine days and \$189 for administrative cost).

\item Price of a mask in South Korea is normally set around \$2 and \$1.2 under the government rationing scheme \cite{amihai_glazer}.
\end{itemize}

\subsection{Lockdown Only Delays the Virus Spread}
\label{subsec:lockdown_fails}
\begin{figure}[h]
    \begin{subfigure}[b]{0.35\linewidth}
    \includegraphics[width=\textwidth]{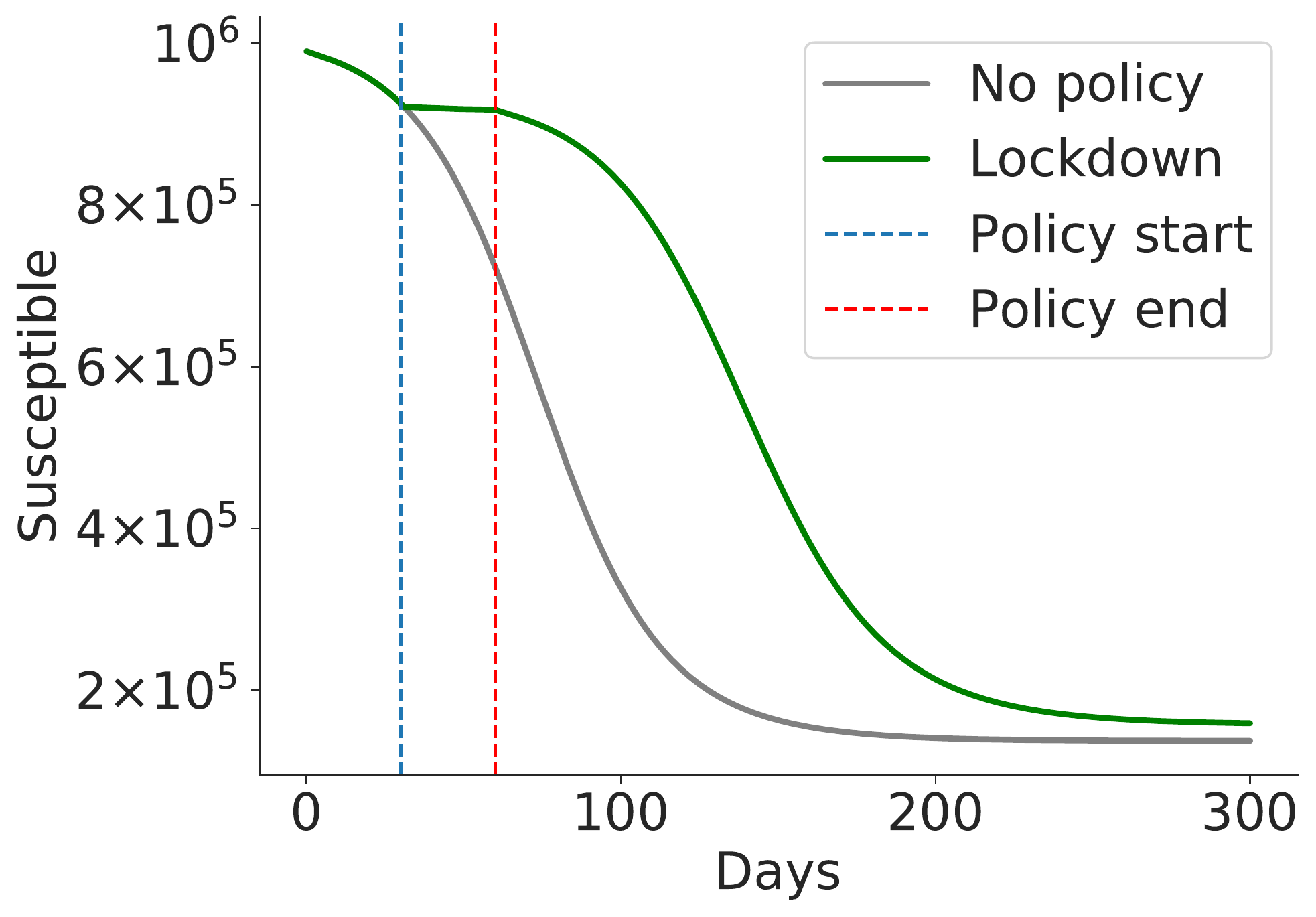}
    \caption{Susceptible cases}
    \end{subfigure} \qquad
    \begin{subfigure}[b]{0.35\linewidth}
    \includegraphics[width=\textwidth]{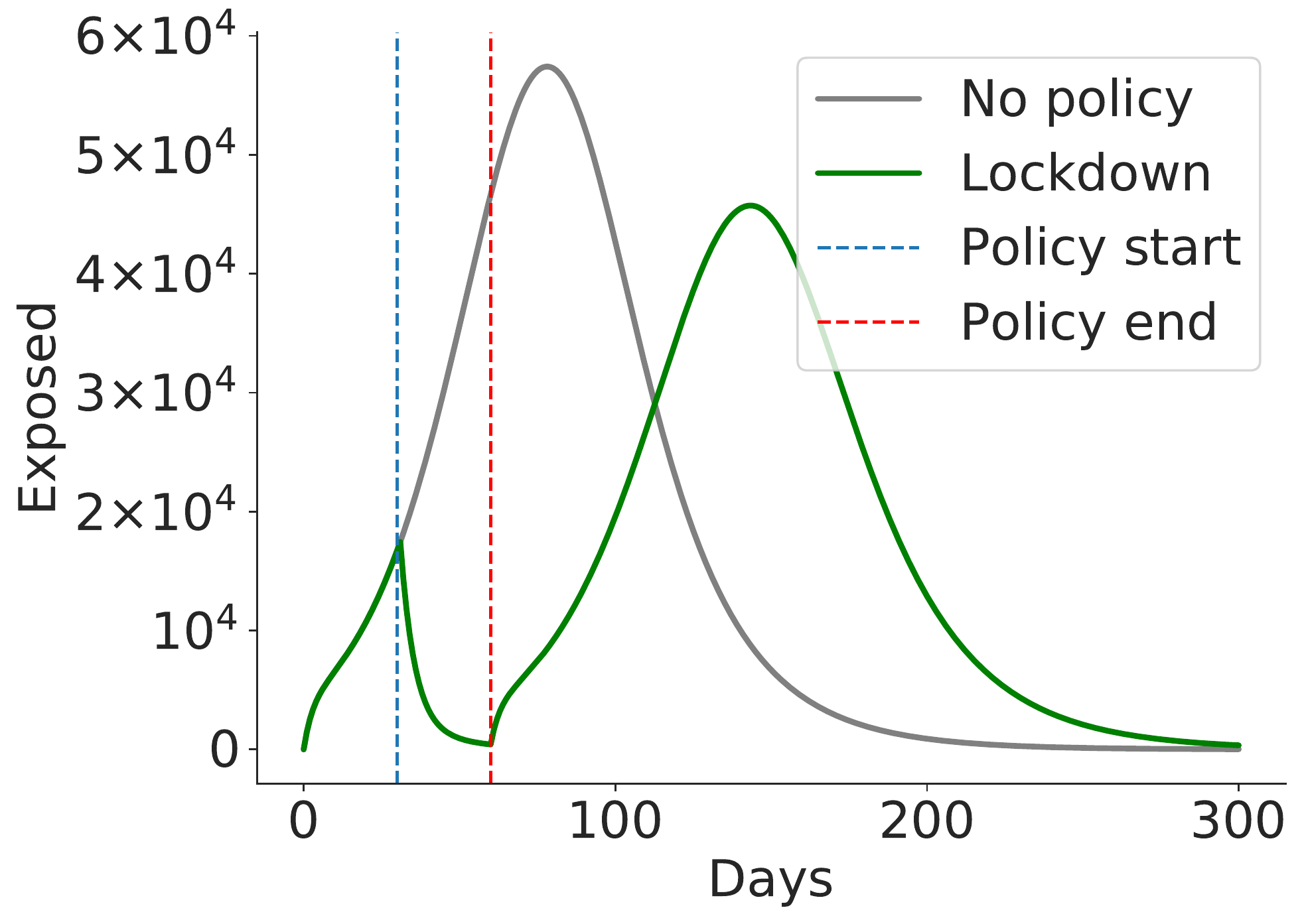}
    \caption{Exposed cases}
    \end{subfigure}
    \begin{subfigure}[b]{0.35\linewidth}
    \includegraphics[width=\textwidth]{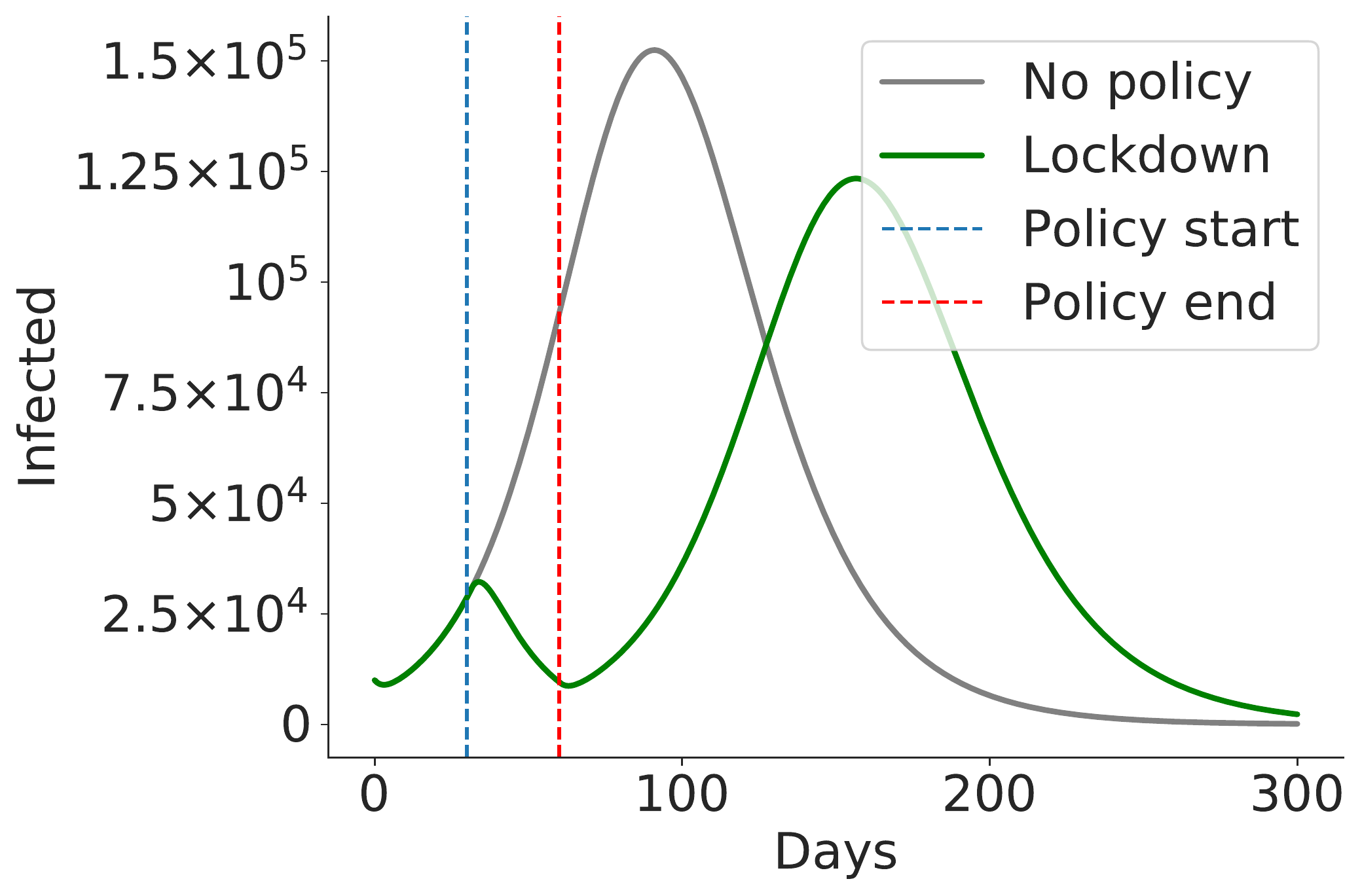}
    \caption{Infected cases}
    \end{subfigure} \qquad
    \begin{subfigure}[b]{0.35\linewidth}
    \includegraphics[width=\textwidth]{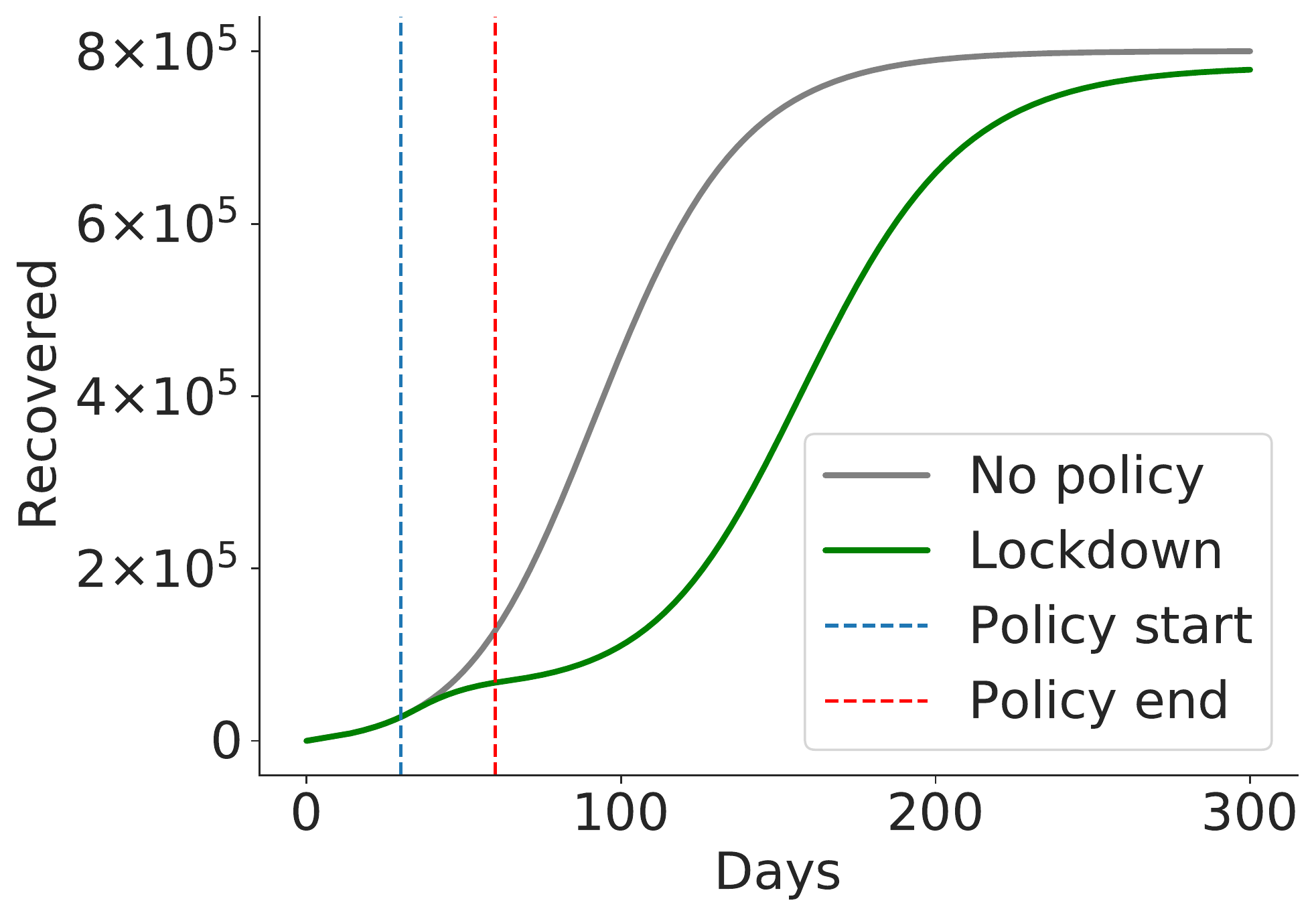}
    \caption{Recovered cases}
    \end{subfigure}
    \begin{subfigure}[b]{0.35\linewidth}
    \includegraphics[width=\textwidth]{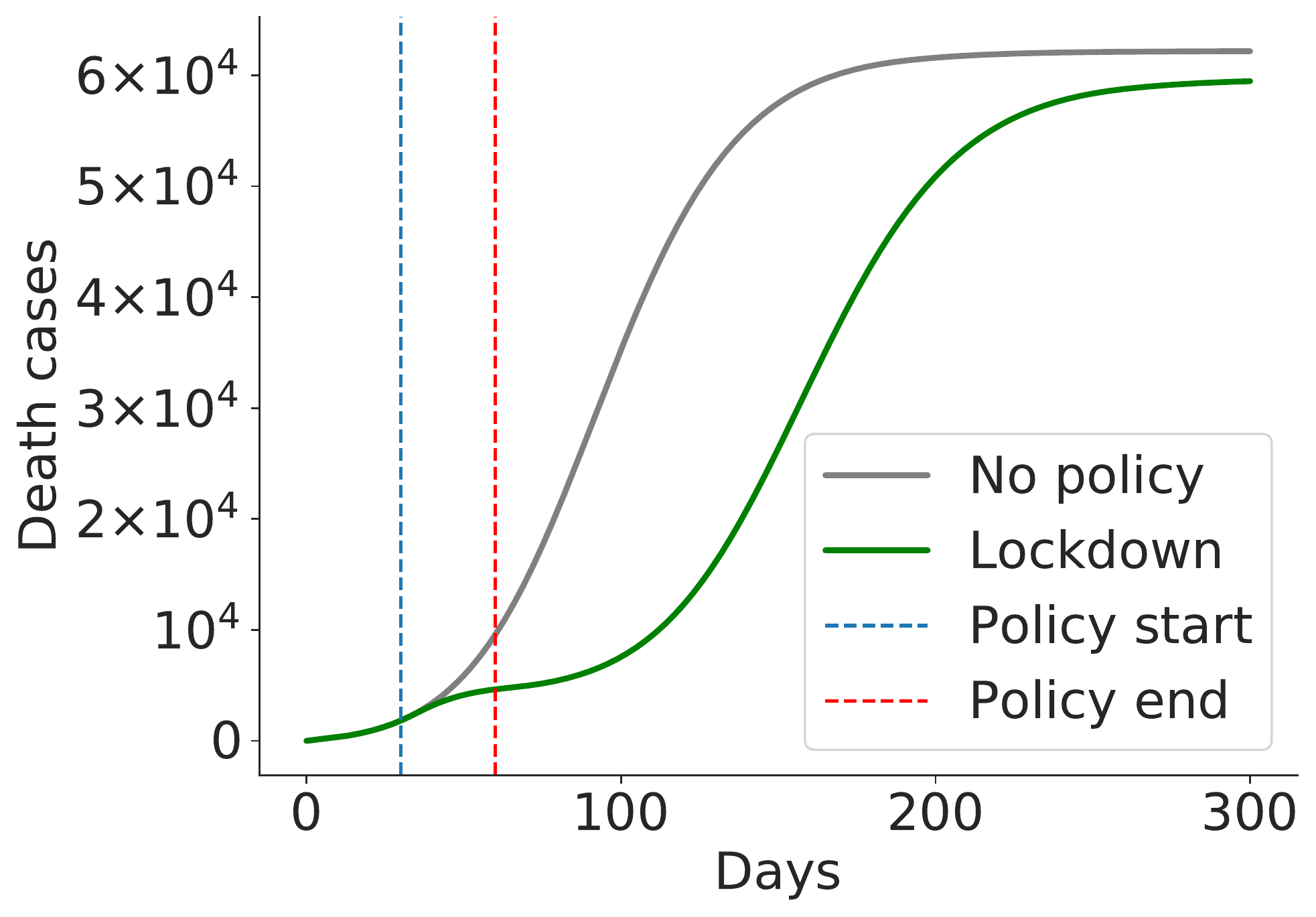}
    \caption{Death cases}
    \end{subfigure} \qquad
    \caption{A lockdown delays the virus spread, but cannot prevent it. In this simulation the lockdown policy was imposed from day 30 to day 60 (total 1 month). The grey graph represents the baseline situation with no policy applied, the green graph indicates the case when a lockdown is applied. The blue and red lines mark the start and end of the lockdown, respectively. We can see that although lockdown sharply decreases the number of exposed and infected cases, it cannot prevent the virus from spreading after the lockdown is lifted on day 60. The number of exposed and infected cases rise again. Given the cost of the lockdown, it is impossible to maintain it for long periods of time, making it less preferable to less costly policies. Thus, conclusion is in alignment with those of prior works \cite{haug2020ranking,sharov2020creating}.}
    \label{ld1}
\end{figure}

We ran the model without any policy and with the lockdown applied from day 30 to 60. As you can see from Fig.~\ref{ld1}, applying lockdown for 1 month simply postpones the virus spread. Another problem is that it significantly hits the country's economy, so it cannot be applied for a long time. Thus, even though lockdown is estimated to have the highest efficiency of 0.96, it might not be the best policy to apply. So further experiments are required to identify how, when, and for how long the policies should be applied.


\subsection {Best Initial Response: Social Distancing With Contact Tracing}
Finally, we want to devise the best initial response to the virus. Using the inferred statistics, we conducted experiments on the policies and performed simulations to develop the optimal policy with minimal loss (both economic loss and life loss). We designed an imaginary country with a population of 1 million and a GDP of \$30,000 per capita. The country had a population of 1 million, and the simulation spanned three months. We assume that policy-makers revised the policy every month, and a policy is applied exhaustively, partially (50\% efficacy), or not applied at all. Policies could be applied together. 
The goal was to minimize the cost. 

\subsubsection{Results}
\input{tables/sim}
\begin{figure}[t]
\centering
\includegraphics[width = 0.7\textwidth]{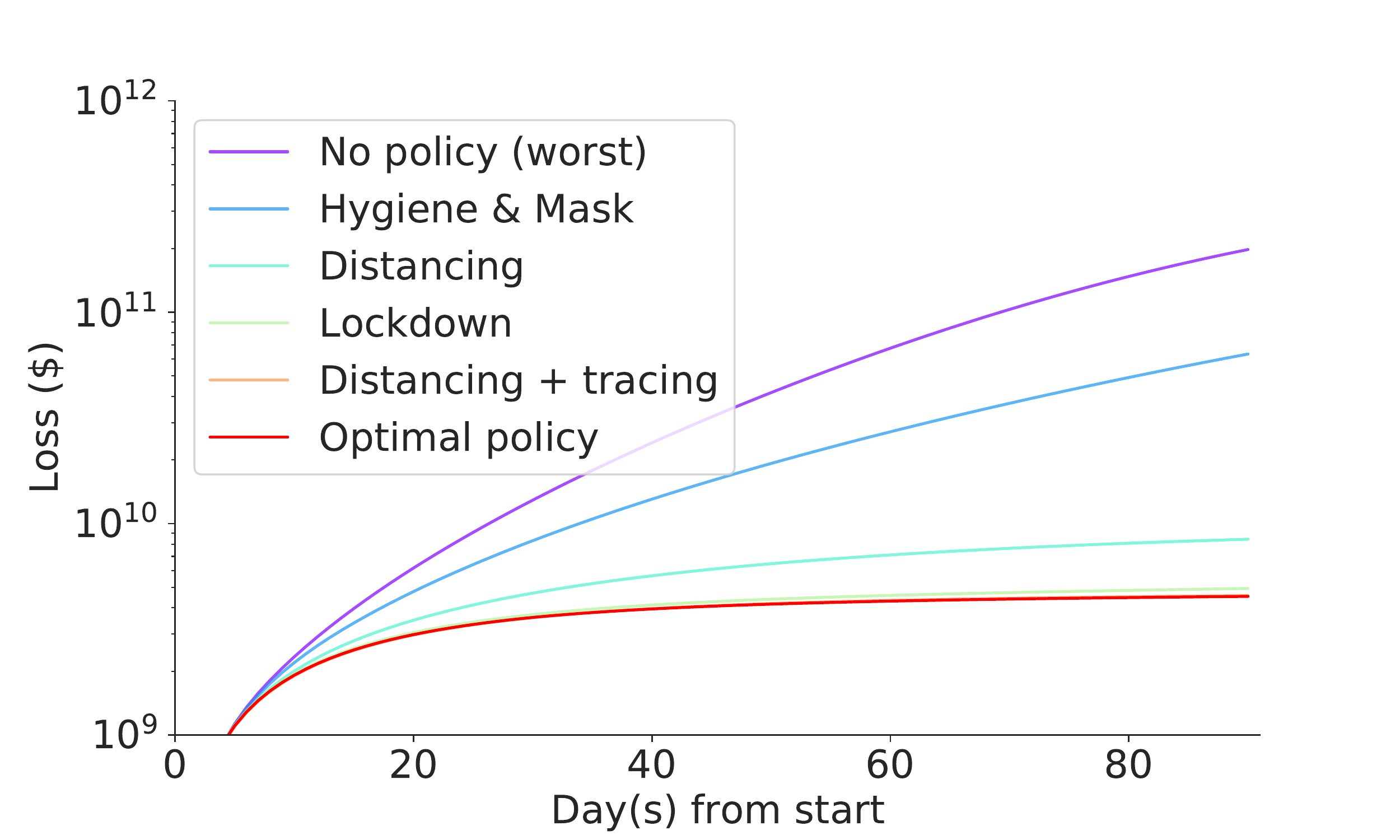}
\caption{The (daily) accumulated loss incurred in each intervention. Social distancing coupled with contact tracing incurred the least loss, followed by lockdown, social distancing, masks and hygiene mandates and no policy incurring the biggest lost.}
\label{loss_graph}
\end{figure}

The after three months results for some important policy combinations are shown in Table \ref{tab:simulation_result_main}.
The full loss trajectories of important policies is shown in Fig.~\ref{loss_graph}.

The best policy identified so far is contact tracing with social distancing, with a loss of around 2 billion dollars. Without intervention, the loss in the imaginary nation is \$197.9 billion. When scaled up to the US scale, this figure reaches \$77 trillion only for the first three months. 

Generally, social distancing coupling with contact tracing incurs less loss than lockdown or social distancing. They are all strong interventions compared to a masks and hygiene mandates. However, they all significantly save a tremendous loss compared to doing nothing or only doing the mask and hygiene mandates. The human cost for mild intervention seems to be significantly lower than the economic cost of the strong intervention. Nonetheless, the masks and hygiene mandates still halved the loss that we suffer when we do nothing.


Contact tracing coupled with social distancing reduces the economic and human capital loss by 98\% compared to doing nothing. Although as efficient as lockdown (\S~\ref{sec:policy}), the economic and human capital costs are at least 8\% less in a 3 month period. The \textit{optimal} policy in our setting is contact tracing and social distancing for three months with additional hygiene and masks mandates for the first month. Hygiene and masks mandates plays some role in minimizing the loss, albeit the improvement is marginal.

Therefore, we can conclude that quarantine and contact tracing are the most efficient policies in our setting. Indeed, we can see that countries that enjoyed the initial success in controlling the virus cases, e.g., South Korea, Vietnam, or Australia applied social distancing and contact tracing as their primary policies.

\subsection{Limitations and Implications}
Due to the challenge of separating the policy effects, our study has some limitations within which our findings need to be interpreted carefully. First, we have investigated major policies applied in relatively wealthy countries. For economic loss estimations, we utilized policies' costs reported from various developed countries using different sources. Depending on the parameters adjusted for a particular country, the results might be different.


We provide a flexible end-to-end pipeline, which could be tailored to each country's specific needs. Given the unique socioeconomic state, the feasible stringency and price tag of the policies vary across the countries. One could adjust the corresponding parameters to adapt to their own situation. 
The resilience to control model's parameters also allows countries to see how the pandemic will play out under different scenarios and build their own strategies based on the model's output. Decision-makers could simulate some hypothetical situations and see the resulting cases, deaths, and capital loss, which assist them in making informed decisions for their citizens.



%% file: tables/sim.tex
\begin{table}[t]
\centering
\caption{Loss regarding some applied policies (more results are available in our repository.)}
\small
\begin{threeparttable}
\begin{tabular}{lrrr}
\toprule
 \textbf{Policy combination} & \textbf{Total Cases}                & \textbf{Total Deaths}              & \textbf{Total Loss (billion \$)}        \\ \midrule
Optimal policy\tnote{*} &  10,734  &  577   &  4.526   \\
 Contact tracing and distancing       &  11,003  &  591   &  4.569  \\
 Lockdown        &  11,003  &  591   &  4.933   \\
 Social distancing        &  22,478  &  1,138   &  8.437   \\
 Mask and hygiene mandate          & 201,929 &  8,941 &  63.400 \\
 No policy      &  592,136 &  28,018 &  197.927 \\ 
\bottomrule
\end{tabular}
\begin{tablenotes}
  \item[*]Optimal policy: Contact tracing and distancing for three months with additional hygiene and mask mandate for the first month.
  \end{tablenotes}
\end{threeparttable}
\label{tab:simulation_result_main}

\end{table}


%% file: 7_conclusion.tex
\section{Conclusion}
\label{sec:conclusion}

Recent research on COVID-19 propagation analysis has provided a deeper understanding of the transmission processes occurring during the past 1.5 years. Epidemiological models point out the key factors that affect the spread of the virus, including the basic reproduction number, virus incubation period, and daily infection number. In the present study, we have moved one step further to gauge the efficacy of the early-stage policy to respond to the pandemic, with economic factors related to the policy itself and its benefits of slowing down the virus. Sophisticated analysis from 10 countries suggests that social distancing, coupled with contact tracing, is the most efficient policy among major interventions. From the data of Asian countries, we derive meaningful results that close contact tracing could provide protection to citizens from the pandemic comparable to lockdowns, without inducing as much cost. Going one step further, we carefully designed a simulated country and gauged the efficacy of each policy combination. Our testbed allows end-users to control various parameters suitable for their country's situation. Through the process of overcoming COVID-19, we are gaining a clearer understanding of the trade-off between virus prevention and economic loss. As we have seen in many countries, it is crucial to identify each policy's efficiencies and costs and to estimate the best time and intensity to impose them before it is too late. We hope that our research will assist every nation in responding to possible future pandemics.